\documentclass[apj]{emulateapj}
\usepackage{epsfig}
\usepackage{amssymb}
\usepackage{amsmath}
\usepackage{graphicx}
\usepackage{latexsym}
\usepackage{dcolumn}
\usepackage{bm}
\usepackage{ulem}
\usepackage{color}
\usepackage{multirow}
\usepackage{cancel}

\newcommand{\bfx}{{\mathbf{x}}}
\newcommand{\dd}{{\mathrm{d}}}

\def\gsim{\;\rlap{\lower 2.5pt
 \hbox{$\sim$}}\raise 1.5pt\hbox{$>$}\;}
\def\lsim{\;\rlap{\lower 2.5pt
   \hbox{$\sim$}}\raise 1.5pt\hbox{$<$}\;}
\def\ie{{\it i.e. }}
\def\eg{{\it e.g. }}

\shorttitle{Testing PSF Interpolation in Weak Lensing with Real Data}
\shortauthors{Lu et al.}

\begin{document}
\title{Testing PSF Interpolation in Weak Lensing with Real Data}

\author{Tianhuan Lu$^1$, Jun Zhang$^{2*}$, Fuyu Dong$^2$, Yingke Li$^2$, Dezi Liu$^3$, Liping Fu$^4$, Guoliang Li$^5$, Zuhui Fan$^3$}
\affil{$^1$Zhiyuan College, Shanghai Jiao Tong University, Shanghai, 200240, China \\
$^2$Department of Physics and Astronomy, Shanghai Jiao Tong University, Shanghai, 200240, China\\
$^3$Department of Astronomy, School of Physics, Peking University, Beijing 100871, China\\
$^4$The Shanghai Key Lab for Astrophysics, Shanghai Normal University, Shanghai  200234, China\\
$^5$Purple Mountain Observatory, Chinese Academy of Sciences, Nanjing, 210000, China
}

\email{*betajzhang@sjtu.edu.cn}

\begin{abstract}

Reconstruction of the point spread function (PSF) is a critical process in weak lensing measurement. We develop a real-data based and galaxy-oriented pipeline to compare the performances of various PSF reconstruction schemes. Making use of a large amount of the CFHTLenS data, the performances of three classes of interpolating schemes -- polynomial, Kriging, and Shepard -- are evaluated. We find that polynomial interpolations with optimal orders and domains perform the best. We quantify the effect of the residual PSF reconstruction error on shear recovery in terms of the multiplicative and additive biases, and their spatial correlations using the shear measurement method of \cite{zhang2015}. We find that the impact of PSF reconstruction uncertainty on the shear-shear correlation can be significantly reduced by cross correlating the shear estimators from different exposures. It takes only 0.2 stars (SNR $\gsim$ 100) per square arcmin on each exposure to reach the best performance of PSF interpolation, a requirement that is satisfied in most of the CFHTlenS data.

\end{abstract}


\section{Introduction}
\label{intro}

Weak lensing (cosmic shear) has been proven to be a powerful tool to reveal the density structure and the expansion history of our Universe \citep{bs01,refregier03,hj08,kilbinger15}. Nevertheless, measuring the weak lensing effect from galaxy shapes turns out to be a highly nontrivial process, mainly due to a number of observational effects, including the PSF effect, the pixelation effect, background noise, Poisson noise, etc.. Historically, the PSF effect is the most well-known and important, as it can distort the galaxy shape to a level that is much larger than that by cosmic shear, and coherent on large scales. Many different methods have been developed for removing the influence of PSF in shear measurement \citep{kaiser95,lk97,hoekstra98,rrg00,kaiser2000,bridle01,bj02,rb03,hs03,mr05,kuijken06,miller07,nb07,kitching08,jz08,ba2014,zhang2015,bernstein2016}.

Removal of the PSF effect relies on accurate reconstruction of the PSF form at the position of the galaxy, which is typically done using neighboring star images. Various interpolating schemes have been proposed, and tested with simulated images \citep{berge2012,gentile2013}. However, some studies have shown that spatial variations of PSF are very difficult to restore \citep{hoekstra2004, jee2011}, due to the atmospheric turbulence and the quality of the star images. The simulated results therefore might not be quite reliable in less-controllable real scenarios. It is desirable to test the accuracy of PSF reconstruction directly using real images.

One can perform this type of tests at the positions of stars, where the PSF form is known. The reconstruction can be done using a different/independent group of stars nearby. The quality of the reconstruction can be checked by directly comparing the recovered PSF image and the original star image, in terms of their sizes, ellipticities, or other properties \citep{hoekstra2004,wittman05,van2005,h12,hamana13}. The residual PSF ellipticity error can be used to estimate the induced shear recovery error through the shear susceptibility factor \citep{paulin2008,rowe2010} in certain shear measurement methods.

We note that different shear measurement methods impose different requirements on PSF reconstruction: some methods only require the size and the ellipticities of the PSF (\eg, \cite{kaiser95}), while most of other more recently developed methods need the whole reconstructed PSF image (see, \eg, \cite{kitching2012,m15}). Therefore, the standard for judging the quality of PSF reconstruction is not quite unique, and its influence on shear measurement is better discussed within a specific shear measurement method. The impact of PSF reconstruction on shear recovery also depends on the galaxy size and morphology. The purpose of this work is to develop a galaxy-oriented pipeline to test PSF reconstruction accuracy for the shear measurement method of \cite{zhang2015} (ZLF15 hereafter).

ZLF15 has a few advantages: 1. it does not require any assumptions on the morphological properties of either the galaxy or the PSF; 2. it contains rigorous ways of removing the systematic errors due to the background noise and the Poisson noise; 3. it is accurate to the second order in shear/convergence; 4. the image processing is very simple and fast ($\sim 0.001$CPU seconds/galaxy), as it only involves Fast Fourier Transformation. In terms of PSF reconstruction, there is another important benefit: ZLF15 only needs the 2D power spectrum of the PSF image, which is automatically centered in Fourier space, avoiding complexities related to sub-pixel level image alignment.

In \S\ref{sec:schemes}, we review three interpolating schemes (polynomial interpolation, Kriging, and Shepard), and the ways they are implemented in the PSF reconstruction of this work. In \S\ref{sec:pipeline}, we lay out the details of the pipeline we use to evaluate the accuracy of PSF reconstruction and its influence on shear recovery, and present our main results achieved using the CFHTlenS data \citep{heymans2012,erben2013}. In \S\ref{sec:discussion}, we discuss the roles of outliers and the overfitting problems in PSF reconstruction, the requirement on stellar number density, and how the properties of our simulated galaxies affect our conclusion. Finally, we conclude in \S\ref{sec:conclusion}.

\section{PSF Interpolation Schemes}
\label{sec:schemes}

\subsection{General Consideration}
\label{sec:general}

The property of PSF that we interpolate in this paper is the 2D power spectrum of the PSF image, as required by ZLF15. This is not equivalent to interpolating the real-space PSF image, as we do not retain the phase information in Fourier space. It is important to note that the image of the 2D power spectrum is automatically centered in Fourier space. This fact greatly simplifies the interpolation process, either on a pixel-by-pixel basis or in terms of principle-component-analysis (PCA) \citep{li2016}. In this work, interpolations are performed on a pixel-by-pixel basis, as we find that PCA does not yield any significant improvement in accuracy\footnote{One should be cautious that this conclusion may only be true regarding the interpolation of the 2D power spectrum of the PSF, since it has no centering/alignment problems.}. The interpolation schemes we consider in this paper include: polynomial fitting, Kriging, and Shepard, which are individually introduced in the rest of this section.

\subsection{Polynomial Fitting}
\label{sec:polynomial}
Polynomial Fitting is a very common and popular scheme in PSF interpolation \citep{van2005,berge2008}. The quantity of interest is modelled as a polynomial function of the spatial position up to a certain order. Polynomials of too low orders are likely to miss some structures of the PSF distributions, and those of too high orders lead to overfitting of the distributions (see \S \ref{sec:overfitting}). The appropriate order of the polynomial function clearly depends on the size of the interpolated region. Regarding the CFHTlenS data, each exposure is about $1^\circ \times 1^\circ$, containing $4 \times 9$ chips of equal sizes. We consider two domain sizes for polynomial fitting: 1. the whole exposure (called global fitting hereafter); 2. each chip (called chipwise fitting hereafter). For global fitting of this paper, we consider four candidate orders: 8, 10, 12, and 14. In the case of chipwise fitting, three orders are considered: 0,1,2. In fitting a polynomial function within a domain, we give every PSF/star power (normalized) equal weightings (as we only use high SNR stars). The 0th-order polynomial fitting therefore simply refers to taking the average of the PSF power spectra within the domain.

\subsection{Kriging Interpolation}
\label{sec:kriging}
Kriging interpolation, originated from geostatistics, has been proposed as an competitive method in PSF interpolation \citep{berge2012}. Kriging is mathematically identical to Gaussian Processes regression, and is strictly valid only when the residuals are normally distributed \citep{ow15}. Kriging interpolation utilizes the variogram of PSF, and interpolates in a way so that the variance of the estimator is minimized \citep{cressie1988}.

\begin{figure}[htbp]
\centering
\includegraphics[width=8cm]{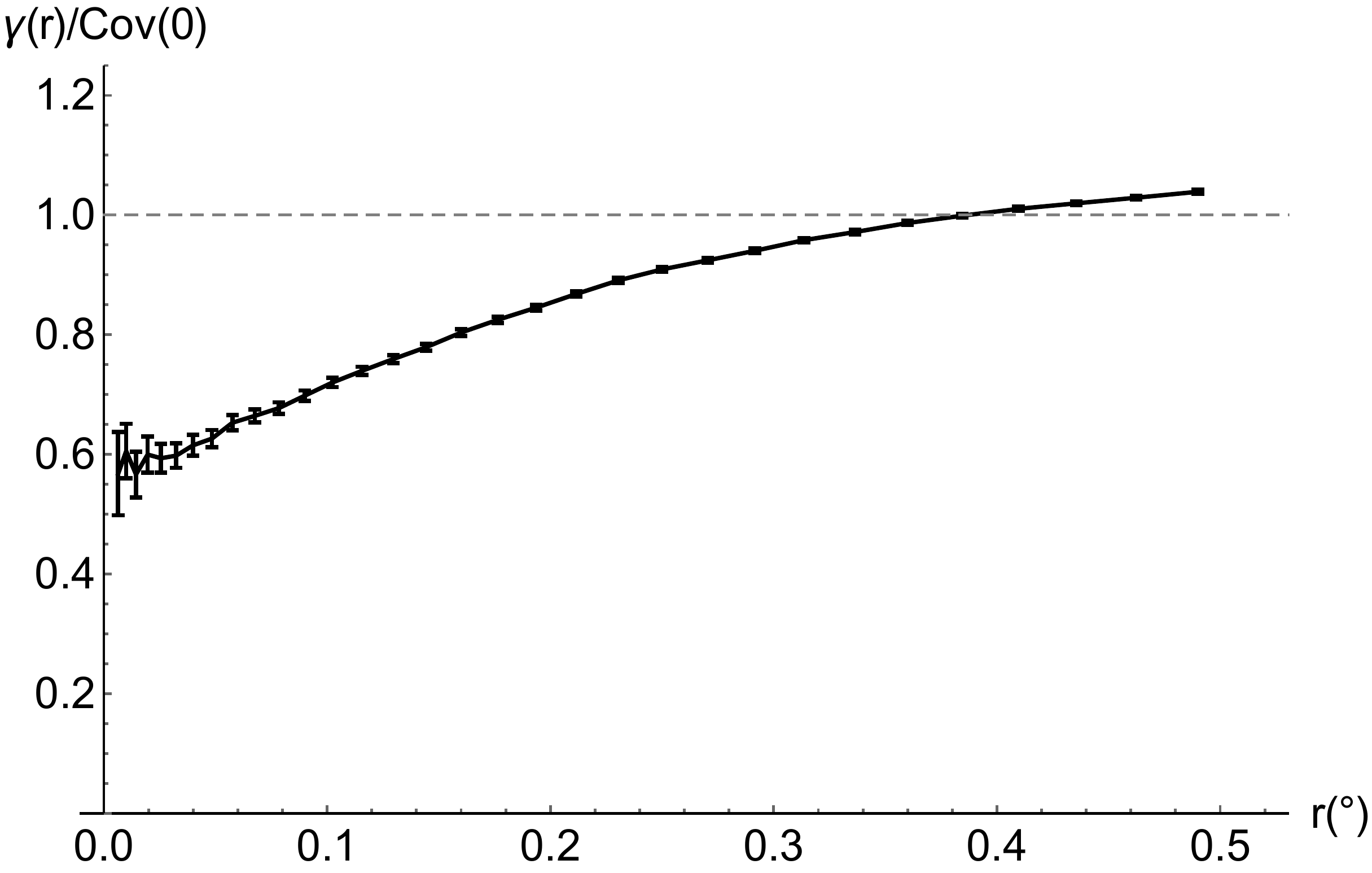}
\caption{The variogram defined in eq.(\ref{gamma}), normalized by $\mathrm{Cov}(0)$. It is calculated using the CFHTlenS data. }
\label{fig:variogram}
\end{figure}

We calculate the variogram of PSF $\gamma$ with the help of the covariance, both of which is a function of spatial separation $\mathbf{r}$, where
\begin{equation}
\label{gamma}
\gamma(\mathbf{r})=\mathrm{Cov}(0)-\mathrm{Cov}(\mathbf{r}).
\end{equation}
We can naturally define $\mathrm{Cov}(\mathbf{r})$ as
\begin{equation}
\mathrm{Cov}(\mathbf{r})=\left\langle \int{(p_{\mathbf{x}}(\mathbf{k})-\bar{p}(\mathbf{k})) (p_{\mathbf{x}+\mathbf{r}}(\mathbf{k})-\bar{p}(\mathbf{k}))  \dd^2 \mathbf{k}} \right\rangle,
\label{eqn:covariance}
\end{equation}
where $p_{\mathbf{x}}(\mathbf{k})$ denotes the normalized PSF power spectrum at the position $\mathbf{x}$, and $\bar{p}(\mathbf{k})$ denotes the average PSF power spectrum within the region of interest. For simplicity, we average $\mathrm{Cov}(\mathbf{r})$ over the direction of $\mathbf{r}$. The variogram based on the CFHTlenS data is shown in Fig.~\ref{fig:variogram}.

Given the star locations $\mathbf{x}_i$ and the corresponding power spectra $p_{\mathbf{x}_i}(\mathbf{k})$, Kriging interpolation estimates $p_{\mathbf{x}}(\mathbf{k})$ at position $\mathbf{x}$ as a weighted linear combination of the PSF power spectra nearby:
\begin{equation}
\label{pweight}
p_{\mathbf{x}}(\mathbf{k})=\sum_{i=1}^{N}w_ip_{\mathbf{x}_i}(\mathbf{k}).
\end{equation}
To meet with the requirements of ``lack of bias'' and ``minimum variance'', the weights are determined by
\begin{equation}
\left[\begin{array}{c}
  w_1    \\
  \vdots \\
  w_N    \\
  \mu
\end{array}\right]
=\left[\begin{array}{cccc}
  \gamma_{11} & \cdots & \gamma_{1N} & 1      \\
  \vdots      & \ddots & \vdots      & \vdots \\
  \gamma_{N1} & \cdots & \gamma_{NN} & 1      \\
  1           & \cdots & 1           & 0
\end{array}\right]^{-1}
\left[\begin{array}{c}
  \gamma_{1\ast} \\
  \vdots         \\
  \gamma_{N\ast} \\
  1
\end{array}\right],
\label{eqn:krigingsys}
\end{equation}
where $\mu$ is the Lagrange multiplier, $\gamma_{ij}=\gamma(|\mathbf{x}_i-\mathbf{x}_j|)$, and $\gamma_{i\ast}=\gamma(|\mathbf{x}_i-\mathbf{x}|)$.

Determining the weights according to Eqn. \eqref{eqn:krigingsys} involves solving a linear system, which can be a quite expensive computation when $N$ is very large. Taking the efficiency into consideration, we only select those PSF that satisfy $|\mathbf{x}_i-\mathbf{x}_\ast|<r_\mathrm{c}$ to interpolate, where $r_\mathrm{c}$ is called the cutoff radius. Note that the variogram almost reaches the maximum and varies very slow at $r \gtrsim 0.3^\circ$, we set $r_\mathrm{c}$ to $0.1^\circ$, $0.2^\circ$, and $0.3^\circ$ in three separate tests to find the balance between performance and efficiency.

\begin{figure}[htbp]
\centering
\includegraphics[width=8cm]{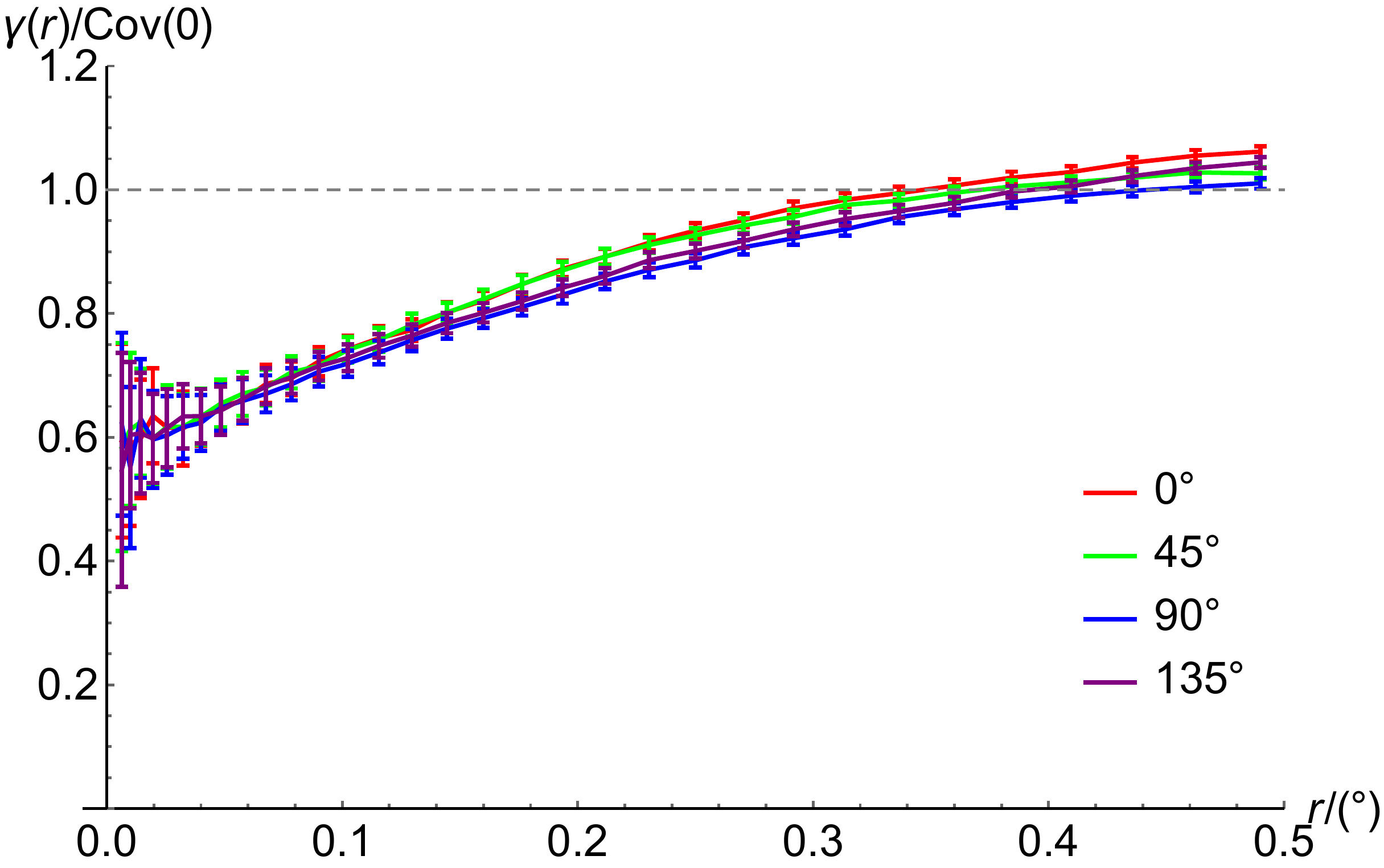}
\caption{The direction-dependent variogram averaged over all the exposures of the W2 field that we use.}
\label{fig:variogram2}
\end{figure}

Finally, we note that the variogram of Kriging can be highly anisotropic due to, \eg, the impact of the wind direction on the atmospheric turbulence \citep{h12}. This anisotropy therefore varies from exposure to exposure. Our variogram is calculated as the average of all the exposures. Its anisotropy is therefore significantly reduced. For example, Fig.~\ref{fig:variogram2} shows the exposure-averaged variograms binned in four different directions. The bin sizes are all 45 degrees. We only use the isotropic variogram in this work, and will consider adopting exposure or direction dependent variogram for the Kriging method in a future work.

\subsection{Shepard Interpolation}
\label{sec:shepard}
Shepard interpolation is a special case of inverse distance weighting (IDW) \citep{shepard68}. Similar to Kriging, Shepard interpolation constructs $p_{\mathbf{x}}(\mathbf{k})$ at $\mathbf{x}$ as a weighted sum of the PSF power spectra from neighboring stars [eq.(\ref{pweight})]. The weighting function takes the form of $w_i=c^{-1}|\mathbf{x}-\mathbf{x}_i|^{-p}$, and $c=\sum_i w_i$.

Unlike Kriging interpolation, which determines weights based on the ``minimum variance'' principal, parameter $p$ in Shepard interpolation cannot be naturally determined. In our tests, three common choices of $p$ are considered: 2, 3, and $\infty$. Larger $p$ means that more weight is put onto the contributions from nearby stars. In the extreme case, $p \rightarrow \infty$ means that the reconstructed PSF is purely determined by its nearest neighbor.

\section{Evaluation Pipeline and Results}
\label{sec:pipeline}

In this section, we present the pipeline that we use to evaluate the performances of different PSF reconstruction schemes. The general idea is to separate the stars into two groups: one is used for constructing PSF; another provides a reference for checking the accuracy of the reconstruction. The shear recovery biases due to PSF reconstruction can be evaluated at positions of the stars of the reference group using simulated galaxies. The spatial correlations of the biases can subsequently be measured to quantify the impact of the PSF uncertainties on the shear-shear correlations. In \S\ref{sec:preprocessing}, we introduce the data source that we use for this study. In \S\ref{sec:reconstruction}, we show the results of PSF reconstruction. The impact of PSF uncertainties on shear recovery is presented in \S\ref{sec:biasmeasuring} and \S\ref{sec:correlationfunction}.

\subsection{Data Source and Preprocessing}
\label{sec:preprocessing}

\begin{figure}[!t]
\centering
\includegraphics[width=8cm]{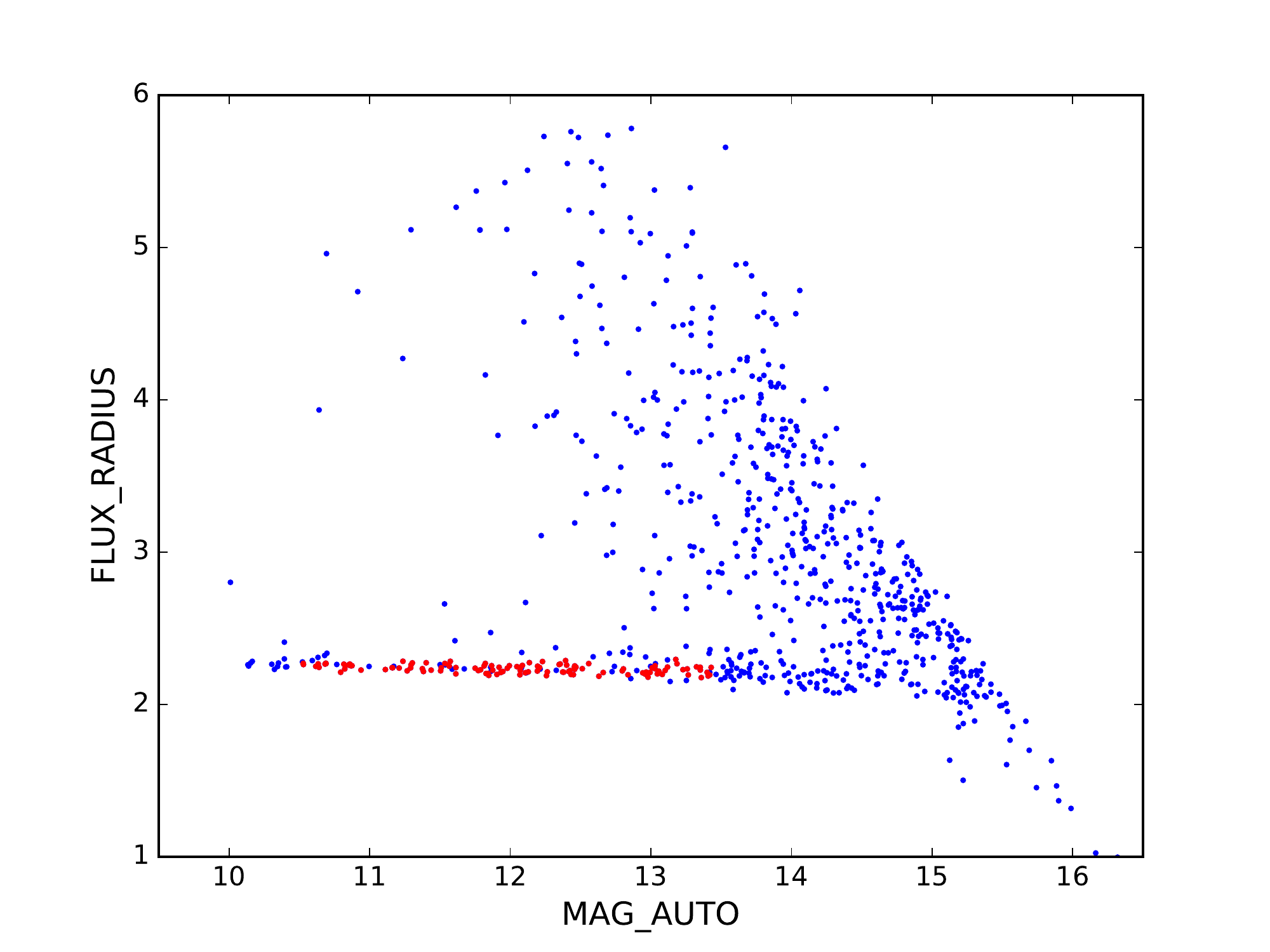}
\caption{ This figure shows how the stars (red points) are selected from the sources (red + blue points) using the Radius-Mag plot (the magnitudes are not calibrated). This is done chip-by-chip. The sources in this figure are from the $20^{th}$ chip of exposure No.831551 of field ``w2m0m0''.}
\label{fig:star_selection}
\end{figure}

The image data we use is from CFHTlenS program \citep{erben2013}. It covers about 150 square degree sky area, separated into four continuous fields (W1, W2, W3, W4). We use the Elixir\footnote{http://www.cfht.hawaii.edu/Instruments/Elixir/} preprocessed CFHTLS-Wide data available at the Canadian Astronomical Data Center (CADC)\footnote{http://www4.cadc-ccda.hia-iha.nrc-cnrc.gc.ca/cadc/}. We use the i'-band single exposures of the W2 field in this study, which has the highest stellar number density among the four fields. The basic preprocessing (background smoothing, cosmic-ray identification, astrometric correction, etc.) of the CCD images are conducted using the THELI software developed by the CFHTlenS team \citep{erben2005,schirmer2013}. The separation of stars from galaxies are done using the well-known 'Radius-Magnitude' plot, which yields little ambiguity in defining the stars, as the stars we use all have signal-to-noise-ratio (SNR) larger than 100. Fig.~\ref{fig:star_selection} shows an example of this type of plot for a typical exposure (the magnitudes in the plot are not calibrated). We remove the stars with their peak fluxes higher than half of the saturation level for avoiding nonlinear CCD effects. Each of our star images is contained in a $48\times 48$ pixel stamp. We run additional routines to remove certain problematic images, such as those contaminated by bad pixels, cosmic rays, neighboring sources (binary stars), etc.. As a result, each chip in the W2 field contains about 100 stars of SNR$>100$, which are used in our PSF reconstruction tests. The results in this paper are achieved using 104 exposures from 16 pointings of the W2 field.

\subsection{PSF reconstuction}
\label{sec:reconstruction}

\begin{figure}[htbp]
\centering
\includegraphics[width=8cm]{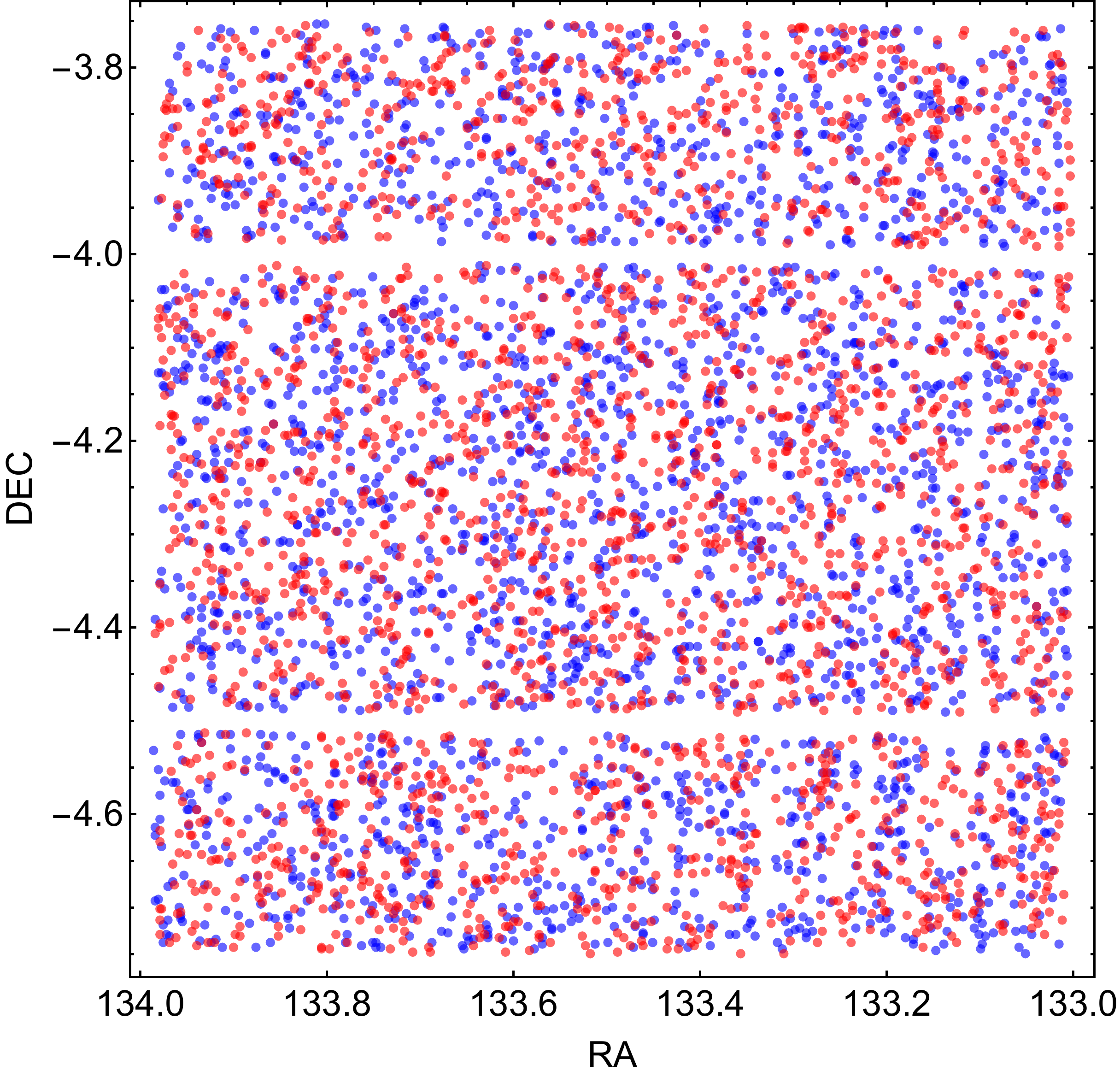}
\caption{The distribution of two groups of stars in one exposure of the w2m0m0 field. The positions of stars in the ``reconstruction'' group and the ``reference'' group are marked by the red and blue circles respectively.}
\label{fig:split}
\end{figure}

The stars in each chip are devided into two groups: one is called the ``reconstruction'' group, which are assigned to the interpolating schemes to recover the PSF distribution; another is the ``reference'' group, for checking the accuracy of the recovery. Regarding the sizes of the two groups, we have the following concerns: 1. the ``reconstruction'' group should be large enough for testing the limitations of interpolation; 2. the size of the ``reference'' group should also be reasonably large, so that we have enough samples to calculate the statistical properties of the PSF reconstruction error. Taking these two factors into consideration, we randomly divide all stars within an exposure into two groups. The resulting two groups have roughly equal sizes and homogenuous star distributions, as shown in Fig.~\ref{fig:split}.

Using the stars in the reconstruction group, we build up the PSF forms at the positions of the reference stars through different interpolation schemes. The quantity that we actually interpolate is the normalized 2D power spectrum of PSF. This is done on a pixel-by-pixel basis. Here normalization means we always set the power at zero wavenumber to be one. The quality of interpolation can be checked by comparing the reconstructed PSF power spectra with those calculated from real stars at all positions of the reference group. The property of the PSF power spectrum we choose to show are the two ellipticities components, which are defined as:
\begin{equation}
\mathbf{e} =\frac{\int{p(\mathbf{k})W(\mathbf{k}) \left(k_x^2 - k_y^2\right) \dd^2 \mathbf{k}}+i\int{2 p(\mathbf{k})W(\mathbf{k}) k_x k_y \dd^2 \mathbf{k}}}{\int{p(\mathbf{k})W(\mathbf{k}) \left(k_x^2 + k_y^2 \right) \dd^2 \mathbf{k}}},
\label{eqn:ellipticity}
\end{equation}
where $W(\mathbf{k})$ is a filter function for reducing the noise contribution at large radii, and defined as:
\begin{equation}
W(\mathbf{k})=\exp\left[-\vert\mathbf{k}\vert^2/(2 \beta^2)\right].
\end{equation}
$\beta$ takes a fix value which is chosen to be larger than the FWHM's of all the stars used.

\begin{figure*}[htbp]
  \centering
  \includegraphics[width=5.3cm]{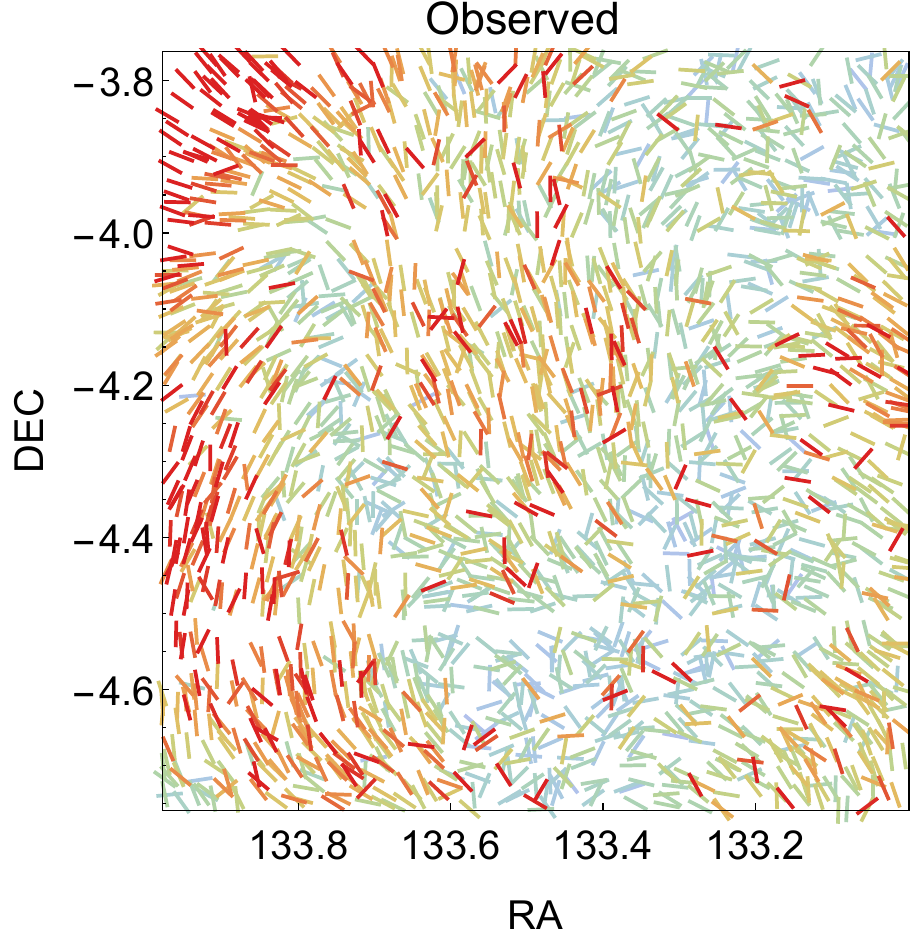}
  \includegraphics[width=11.5cm]{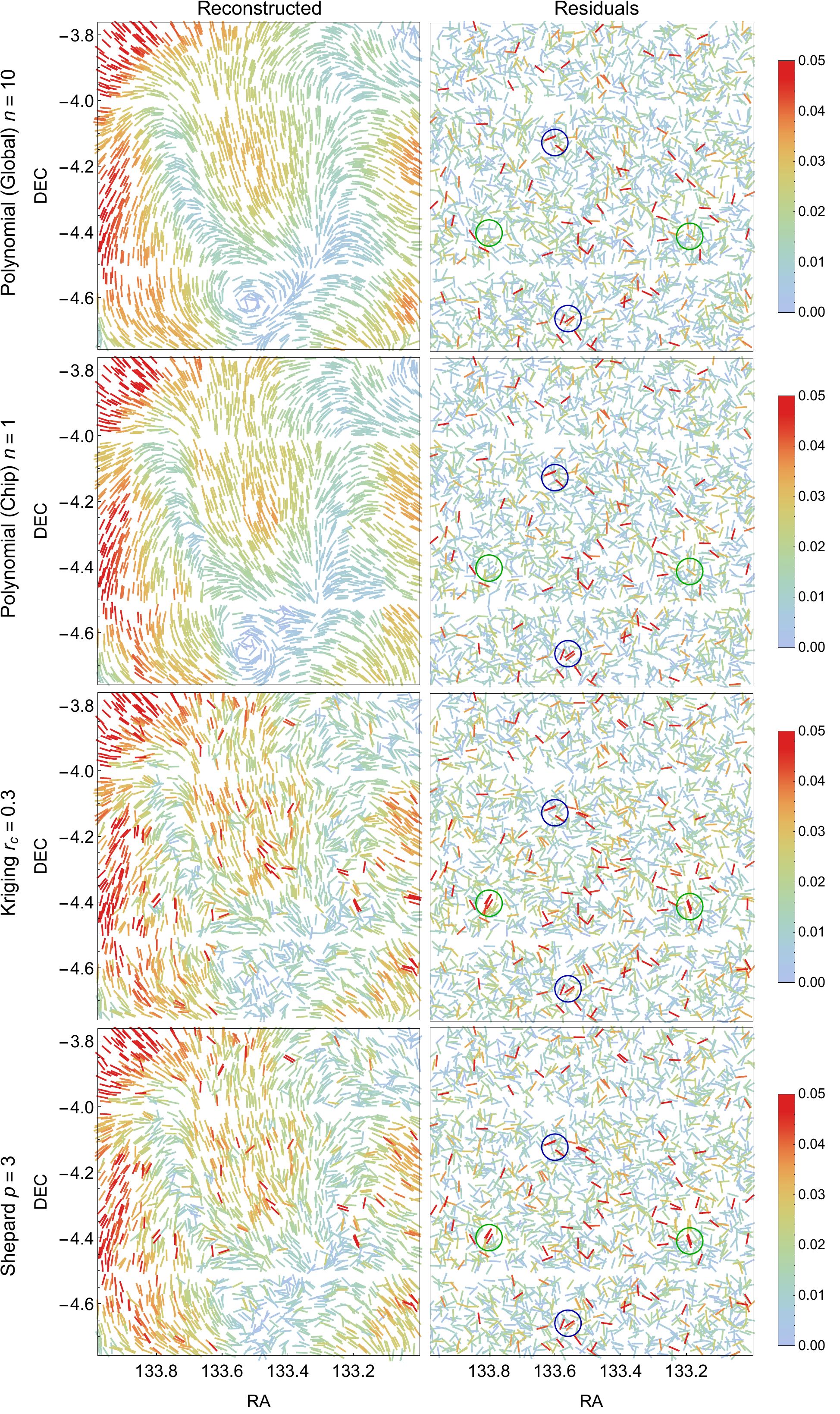}
  \caption{The observed, reconstructed, and residual fields of the PSF ellipticities on a randomly chosen exposure of w2m0m0. The magnitudes of the ellipticities are indicated by color. In the figures of the rightmost column, we use blue circles to mark exemplary places where the PSF ellipticities cannot be accurately reconstructed by all schemes, and green circles to show places where Kriging and Shepard seem to be affected by some extraordinary interpolating stars, while polynomial implementations do not. }
  \label{fig:residualplot}
\end{figure*}

In Fig.~\ref{fig:residualplot}, we show the comparison of the reconstructed PSF ellipticities with those from the reference stars for the following interpolation schemes:
\begin{itemize}
  \item global polynomial of 10th order;
  \item chipwise polynomial of 1st order;
  \item Shepard with $p = 3$;
  \item Kriging with $r_\mathrm{c} = 0.3^\circ$.
\end{itemize}
From the left to the right columns, we present the observed, the reconstructed, and the residual fields of ellipticities respectively on a randomly chosen exposure of w2m0m0. Generally, extremely large residuals occur more frequently in Kriging and Shepard than in two polynomial implementations. Star-by-star comparison shows that there are certain stars (marked by blue circles) that cannot be accurately reconstructed by all schemes, and there are some regions (marked by green circles) where Kriging and Shepard seem to be affected by some extraordinary interpolating stars while polynomial implementations do not. It is observed that in green circles, residuals of Kriging pointing to similar directions, contributing to short-range correlations.

\subsection{Induced Shear Recovery Bias}
\label{sec:biasmeasuring}

The shear recovery bias due to the PSF reconstruction uncertainty can be checked at the position of each reference star with the following procedures:
\begin{itemize}
  \item Generate a large number of galaxy images, and distort each of them by a random shear $(g_1,g_2)$ as input;
  \item Convolve the images with the observed PSF given directly by the reference star;
  \item Measure the shear output $(\tilde{g}_1,\tilde{g}_2)$ using the PSF interpolated from the stars in the ``reconstruction'' group;
  \item Find the multiplicative and additive biases $m_i$ and $c_i$ (i=1,2) by fitting linear relations between $g_i$ and $\tilde{g}_i$: $\tilde{g}_i = (1 + m_i) g_i + c_i \; (i=1,2)$.
\end{itemize}
The resulting spatial and numerical distribution of the shear biases $m_i$ and $c_i$ can be further used for studying the average impact of PSF reconstruction on shear recovery and shear-shear correlations. Note that in principle, this pipeline can work with any shear measurement method.

In our tests, each galaxy is made of 20 point sources of equal luminosities, distributed according to two-dimensional Gaussian distribution. Using the PSF form given directly by the reference star, we generate the power spectrum of the galaxy image in Fourier space, in which each point source simply contributes a plane-wave. The power spectrum of the galaxy and the reconstructed PSF are then used for shear recovery with the method of ZLF15. No noise is added in our tests. The average size of these simulated galaxies is similar to the size of the PSF, whose FWHM is about $3\ \mathrm{pixel}$. The stamp size in our simulations is $48\ \mathrm{pixel} \times 48\ \mathrm{pixel}$.

Note that a galaxy image generated by random point sources is not isotropic, mimicking the so-called ``shape noise''. Shape noise does not affect the expectation of biases, but leads to a large uncertainty. To reduce shape noise, we use each image repeatedly with four different rotations -- $0^\circ$, $45^\circ$, $90^\circ$, and $135^\circ$, and take the average of four $\tilde{g}_i$ as the measurement. Image rotations can be easily done on point sources. The input cosmic shears $g_1$ and $g_2$ are both generated according to Gaussian distribution for each galaxy image independently, with rms of about $0.01$, which is similar to the amplitude of cosmic shear.

\begin{figure}[htbp]
\centering
\includegraphics[width=8cm]{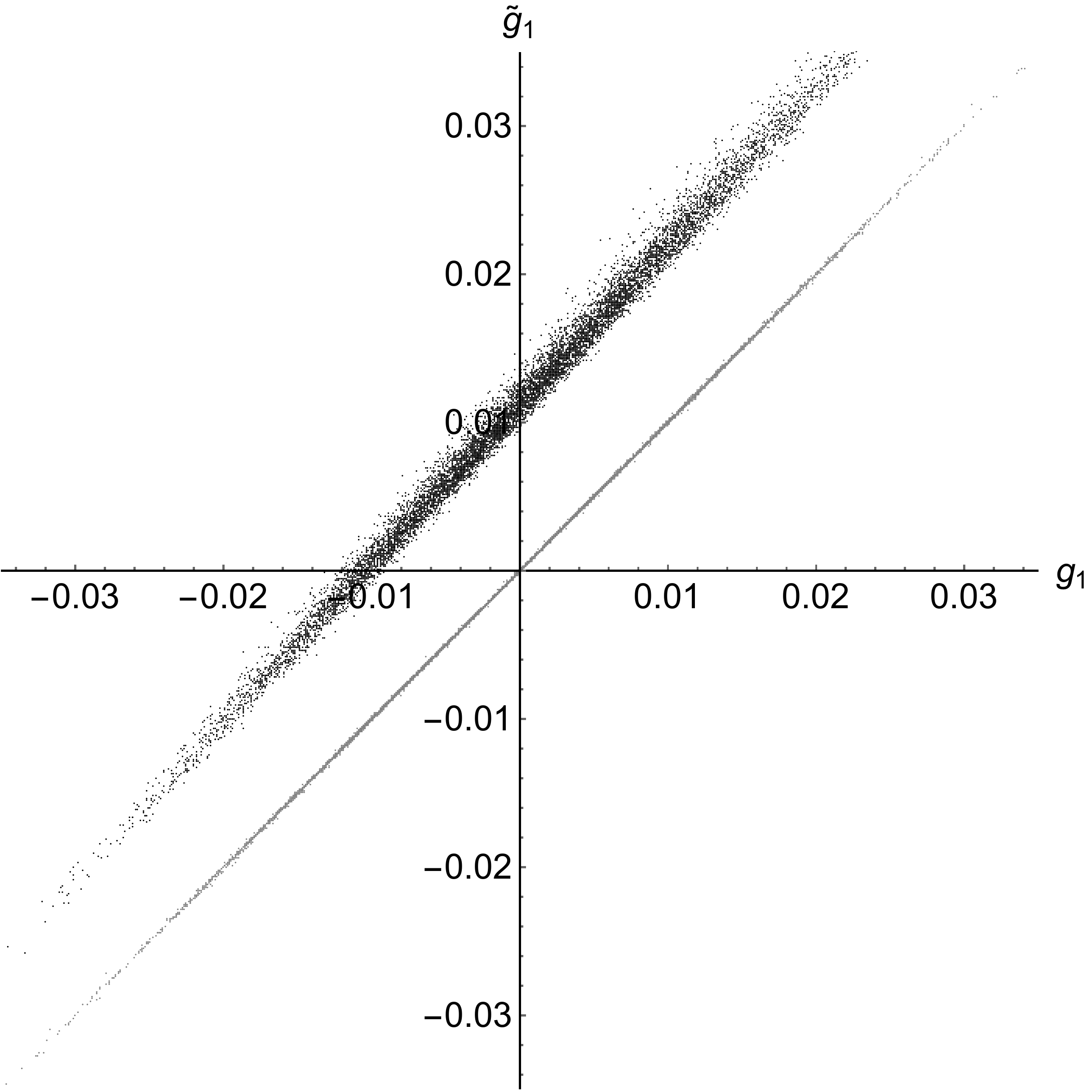}
\caption{The recovered shears are plotted against the input shear values at the position of one typical reference star. The relationship between $g_i$ and $\tilde{g}_i$ is linear. The best-fit values for the multiplicative and additive biases are $m_1=0.072 \pm 0.001$ and $c_1=0.01132 \pm 0.00001$ (from the more dispersed data points). Ten thousand data points are used in this example. The more concentrated data points are from our reference case, with $m_1=(1.15\pm0.06)\times10^{-3}$, $c_1=(-0.7\pm0.6)\times10^{-6}$. This is achieved using the star on-site as the PSF, therefore does not contain any PSF error.}
\label{fig:biasfit}
\end{figure}

Fig.~\ref{fig:biasfit} shows the recovered shears plotted against the input shear values at the position of one typical reference star. In our tests, 10 thousand data points are used on each site to calculate the multiplicative and additive biases $(m_i,c_i)$. The more dispersed distribution in the figure is a typical case using the reconstructed PSF, corresponding to $m_1=0.072 \pm 0.001$ and $c_1=0.01132 \pm 0.00001$. Also shown as a reference is the much more concentrated distribution of data points that passes through the origin of the plot. It is achieved using the star on-site as the PSF, yielding $m_1=(1.15\pm0.06)\times10^{-3}$, $c_1=(-0.7\pm0.6)\times10^{-6}$. Note that the reference case guarantees that the $m$ and $c$ derived using the reconstructed PSF are purely caused by the PSF error, \ie, the morphological differences between the reference star and the reconstructed PSF. This fact is due to the high accuracy and model independence of ZLF15.

\begin{figure*}[htbp]
\centering
\includegraphics[width=17cm]{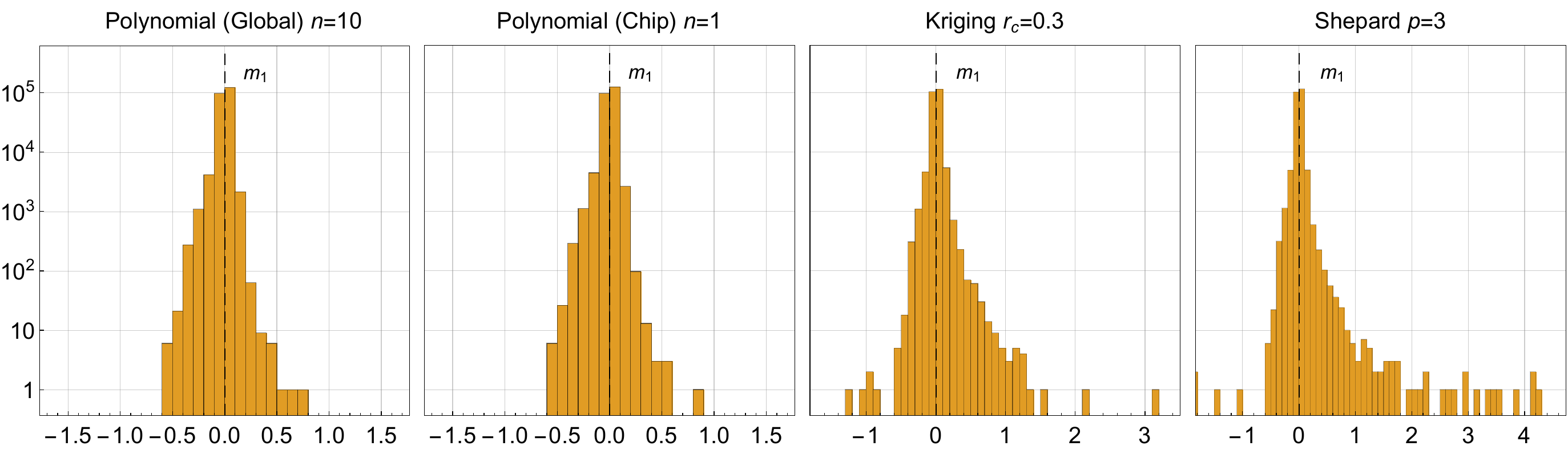}
\includegraphics[width=17cm]{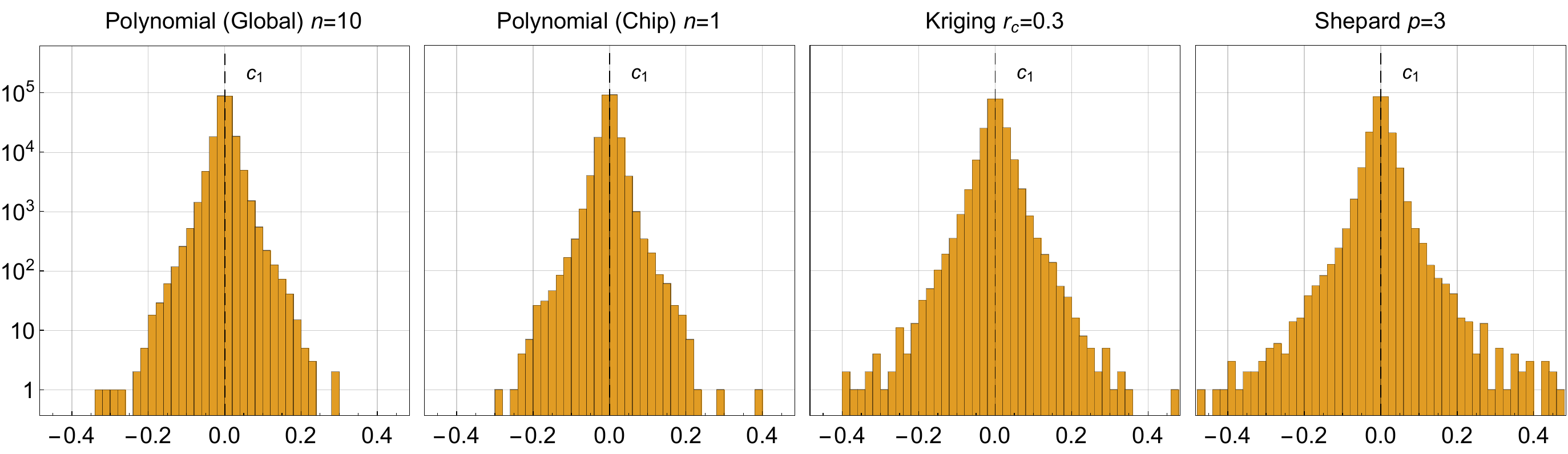}
\includegraphics[width=17cm]{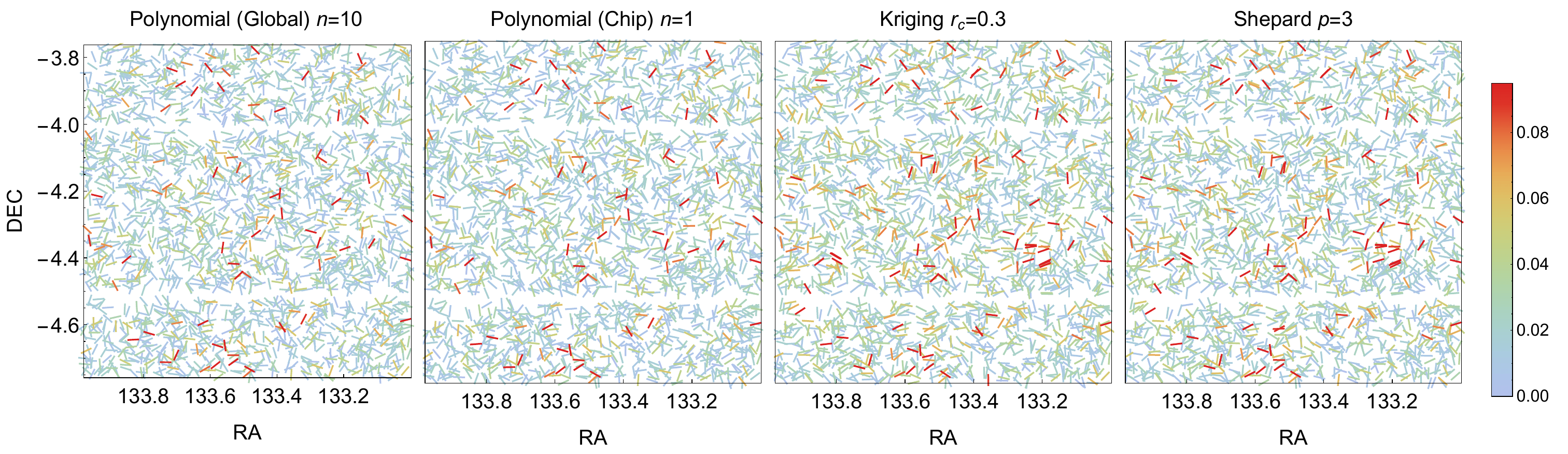}
\caption{The upper eight panels show the numerical distributions of $m_1$ and $c_1$. The lower four panels show the spatial distributions of ($c_1,c_2$). The color in the lowest panel indicates the amplitude of $\sqrt{c_1^2+c_2^2}$.}
\label{fig:dist}
\end{figure*}

An optimal PSF reconstruction method should keep the following quantities sufficiently small: 1. the means of the biases; 2. the deviations of the biases. The upper two panels of Fig.~\ref{fig:dist} show the numerical distributions of $m_1$ and $c_1$ respectively, and the lowest panel shows the spatial distribution of $(c_1,c_2)$ under four interpolation methods. Qualitatively, one already tends to conclude that the two polynomial methods work better than the Kriging or Shepard methods, as the later two both contain long tails in the distributions of $m$.

More quantitatively, we can compare the means and deviations of $m$ and $c$ defined, \eg, as:
\begin{eqnarray}
\langle m \rangle &=& \frac{1}{2} \left( \langle m_1 \rangle + \langle m_2 \rangle \right), \\
\langle c \rangle &=& \frac{1}{2} \left( \langle c_1 \rangle + \langle c_2 \rangle \right),
\end{eqnarray}
and define the deviations of biases as
\begin{eqnarray}
{\langle m^2 \rangle}^{1/2} &=& \frac{1}{2} \left( {\langle m_1^2 \rangle}^{1/2} + {\langle m_2^2 \rangle}^{1/2} \right), \\
{\langle c^2 \rangle}^{1/2} &=& \frac{1}{2} \left( {\langle c_1^2 \rangle}^{1/2} + {\langle c_2^2 \rangle}^{1/2} \right).
\end{eqnarray}

According to the results shown in Table \ref{tab:mean}, we find that Kriging and Shepard are generally worse than the polynomial fitting methods in terms of $\langle m \rangle$ and ${\langle m^2 \rangle}^{1/2}$. In addition, the chipwise polynomial fitting with $n=0$ is obviously worse than the other polynomial methods in terms of $\langle m \rangle$ and $\langle c \rangle$. 

\begin{table*}[htbp]
\centering
\caption{Means and Deviations of Shear Biases}

\begin{tabular}{c|c||c|c|c|c}
\hline \hline
Scheme & Implementation &
  $\langle m \rangle / 10^{-4}$ &
  ${\langle m^2 \rangle}^{1/2} / 10^{-2}$ &
  $\langle c \rangle / 10^{-5}$ &
  ${\langle c^2 \rangle}^{1/2} / 10^{-2}$ \\
\hline
\multirow{4}{*}{Polynomial}
& Global, n=8  & $3.1\pm0.7$ & $4.5$ & $-1.3\pm3.0$ & $2.1$ \\
& Global, n=10 & $3.8\pm0.7$ & $4.5$ & $-0.3\pm3.0$ & $2.1$ \\
& Global, n=12 & $4.3\pm0.7$ & $4.5$ & $0.4\pm3.0$ & $2.1$ \\
& Global, n=14 & $4.8\pm0.7$ & $4.6$ & $0.8\pm3.0$ & $2.1$ \\
\hline
\multirow{3}{*}{Polynomial}
& Chip,   n=0  & $31.2\pm0.8$ & $5.5$ & $36.5\pm3.4$ & $2.3$ \\
& Chip,   n=1  & $8.3\pm0.7$ & $4.5$ & $6.9\pm3.0$ & $2.1$ \\
& Chip,   n=2  & $8.3\pm0.7$ & $4.5$ & $1.3\pm3.0$ & $2.1$ \\
\hline
\multirow{4}{*}{Kriging}
& $r_\mathrm{c}=0.1^\circ$
               & $27.4\pm1.3$ & $9.0$ & $1.0\pm3.8$ & $2.6$ \\
& $r_\mathrm{c}=0.2^\circ$
               & $29.8\pm1.6$ & $10.9$ & $2.2\pm3.8$ & $2.6$ \\
& $r_\mathrm{c}=0.3^\circ$
               & $28.8\pm1.5$ & $9.9$ & $1.3\pm3.8$ & $2.6$ \\
\hline
\multirow{3}{*}{Shepard}
& p=2          & $-3.6\pm1.1$ & $7.6$ & $-1.9\pm3.3$ & $2.2$ \\
& p=3          & $20.0\pm2.8$ & $18.7$ & $-4.3\pm4.9$ & $3.3$ \\
& p=$\infty$   & $74.7\pm6.9$ & $46.7$ & $9.8\pm11.1$ & $7.5$ \\
\hline \hline
\end{tabular}

\label{tab:mean}
\end{table*}

\subsection{Spatial Correlations of the Biases}
\label{sec:correlationfunction}

Given the spatial distribution of the shear biases $m$ and $c$, one can subsequently study their correlation functions, which have direct impacts on the shear-shear correlations. For convenience, let us first consider the correlation between each shear component, which is defined as:
\begin{equation}
C_{g_i}(\theta)=\langle g_i(\bfx) g_i(\bfx + \theta)\rangle \; (i=1,2).
\end{equation}
Due to the existence of biases, the measured correlation function can be written as:
\begin{eqnarray}
\nonumber C_{\tilde{g}_i}(\theta)
    &=& \langle \tilde{g}_i(\bfx) \tilde{g}_i(\bfx + \theta)\rangle \\
    &=& C_{c_i}(\theta) + \left( 1 + 2\langle m_i \rangle + C_{m_i}(\theta) \right) C_{g_i}(\theta),
\end{eqnarray}
where
\begin{eqnarray}
\label{eqn:selfcorrelation}
C_{m_i}(\theta) &=& \langle m_i(\bfx) m_i(\bfx + \theta)\rangle \\
C_{c_i}(\theta) &=& \langle c_i(\bfx) c_i(\bfx + \theta)\rangle.
\end{eqnarray}

The accuracy of the measurements of $C_{g_i}$ are related to the magnitude of $C_{m_i}$ and $C_{c_i}$. In our tests, we observe that $C_{m_1}\approx C_{m_2}$ and $C_{c_1}\approx C_{c_2}$. Therefore, in terms of order-of-magnitude estimates, we find that it is enough to use the following quantities ($C_m$ and $C_c$) to demonstrate the performances of interpolating schemes on correlation functions:

\begin{eqnarray}
\label{eqn:correlationtotal}
C_m(\theta) &=& C_{m_1}(\theta) + C_{m_2}(\theta), \\
C_c(\theta) &=& C_{c_1}(\theta) + C_{c_2}(\theta).
\end{eqnarray}
Note that $C_c(\theta)$ directly corresponds to the correction to the usual ``+'' component ($\xi_+$) of the shear-shear correlation function defined in the coordinates with the x-axis connecting two galaxies (see, \eg, \cite{kilbinger2013}).

Since we have shear measurement on each exposure, we can choose to study the correlation between the shear biases on the same exposure (self correlation) or on different exposures (cross correlation). The importance of these two kinds of tests depends on the shear measurement strategy: in the shear-shear correlation measurement, whether the pair of shear estimators are derived from the same or different exposures. In most shear measurement methods, we find that information from different exposures are all involved in calculating the shear estimators of a galaxy. In this case, it is not very informative to discuss these two kinds of tests separately. However, in principle, all shear measurement methods allow the calculation of shear estimators on single exposures individually. In general, it is not clear (see \cite{bj02}) what the best way is for combining shape/shear information from different exposures. At least in the method of ZLF15, both options are valid: one can either directly correlate the shear estimators derived from either the same exposure or different exposures, or one can combine the shear estimators from different exposures for each galaxy first, and then count on the galaxy ID's for shear-shear correlation measurement. We therefore think it is useful to study both the self and cross correlations of the shear biases.

Interestingly, we find that {\it cross-correlating the shear estimators derived from different exposures can significantly reduce the correlation of shear biases}. It echoes the suggestions of, \eg, \cite{jj04} and \cite{h12}. We can define such correlations as:
\begin{equation}
C_{\tilde{g}_i}^{kl}(\theta)=\langle \tilde{g}_i^{(k)}(\bfx) \tilde{g}_i^{(l)}(\bfx + \theta)\rangle,
\end{equation}
where $k$ and $l$ stand for the labels of the exposures. The relation between $C_{\tilde{g}}^{kl}(\theta)$ and $C_g(\theta)$ is:
\begin{equation}
C_{\tilde{g}_i}^{kl}(\theta)= C_{c_i}^{kl}(\theta) + \left( 1 + 2 \langle m_i \rangle + C_{m_i}^{kl}(\theta) \right) C_{g_i}(\theta),
\label{eqn:correlationdistortion}
\end{equation}
where
\begin{eqnarray}
C_{m_i}^{kl}(\theta) &=& \langle m_i^{(k)}(\bfx) m_i^{(l)}(\bfx + \theta)\rangle \\
C_{c_i}^{kl}(\theta) &=& \langle c_i^{(k)}(\bfx) c_i^{(l)}(\bfx + \theta)\rangle.
\end{eqnarray}
Note that if we define the correlation function of shear as:
\begin{equation}
C_{\tilde{g}_i}(\theta)=\frac{2}{n(n-1)}\sum\limits_{k<l}{C_{\tilde{g}_i}^{kl}(\theta)},
\end{equation}
where $n$ denotes the number of available exposures (which is 6 or 7 in our tests depending on the pointing), eq.\eqref{eqn:selfcorrelation} should be rewritten as:
\begin{eqnarray}
\label{eqn:crosscorrelation}
C_{m_i}(\theta) &=& \frac{2}{n(n-1)}\sum\limits_{k<l}{C_{m_i}^{kl}(\theta)}, \\
C_{c_i}(\theta) &=& \frac{2}{n(n-1)}\sum\limits_{k<l}{C_{c_i}^{kl}(\theta)}.
\end{eqnarray}

\begin{figure}[htbp]
  \includegraphics[width=8cm]{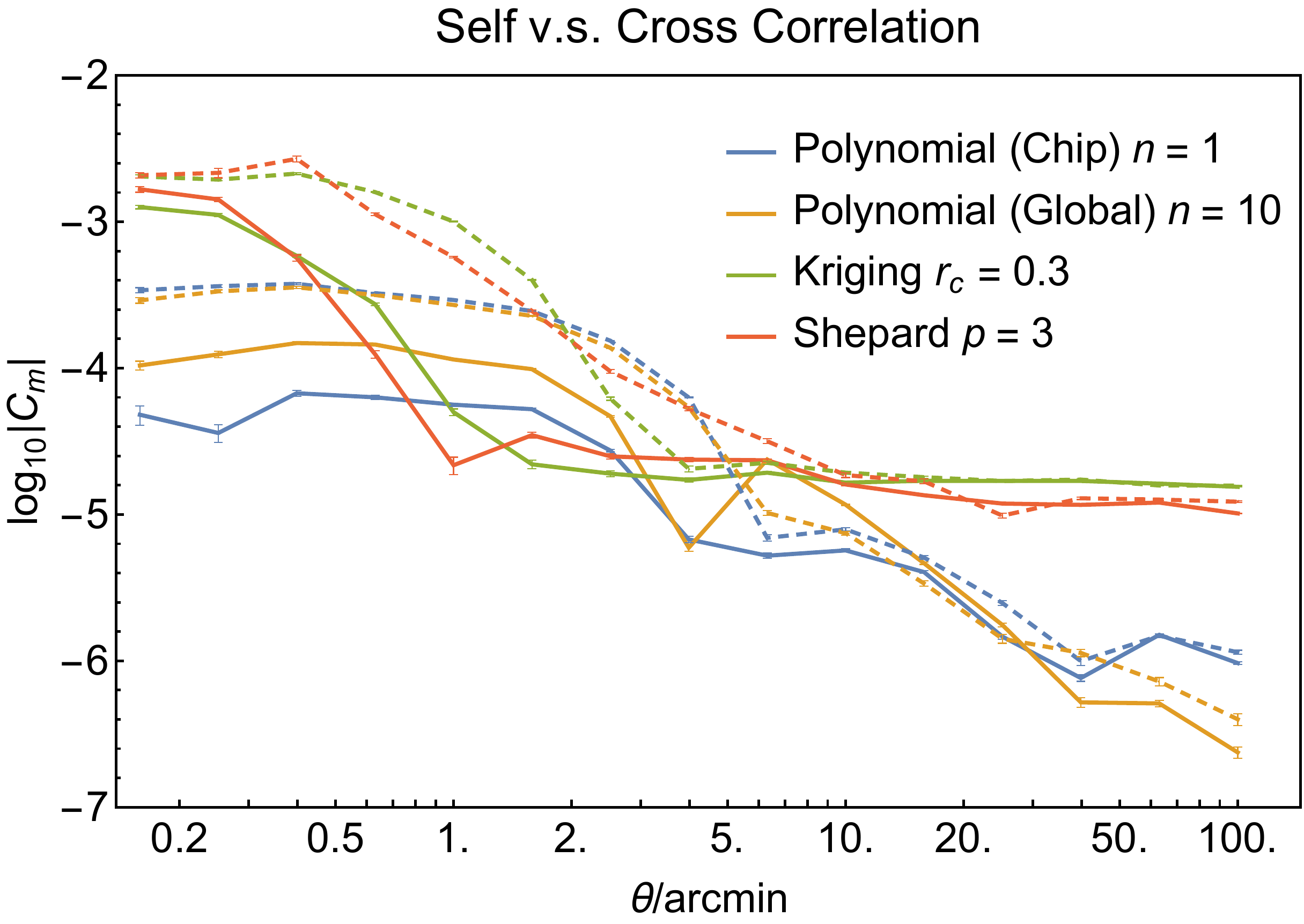}
  \includegraphics[width=8cm]{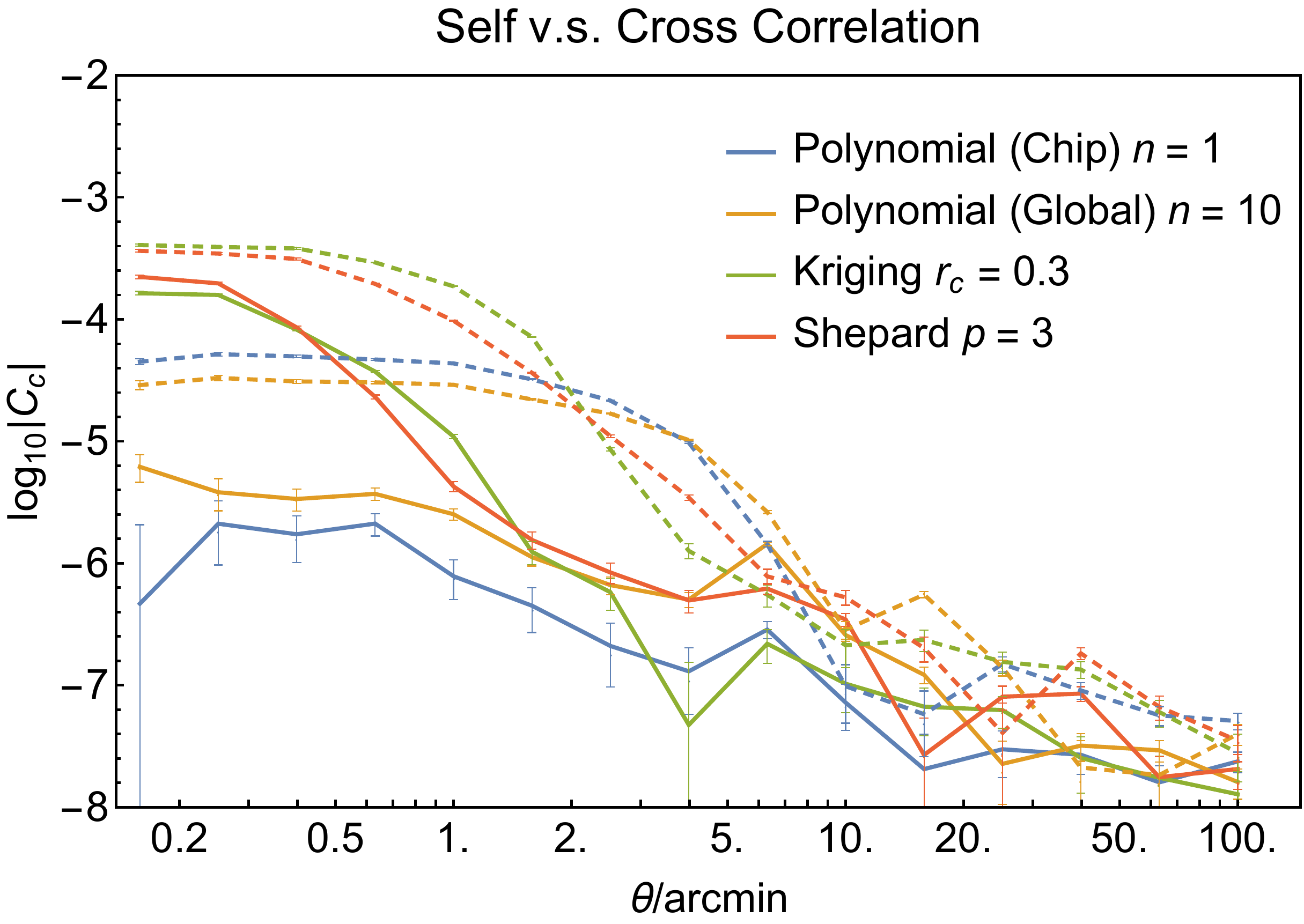}
  \caption{The spatial correlation functions of the additive and multiplicative biases calculated from the same exposure (self-correlation, marked by the dashed curves) and different exposures (cross-correlation, marked by the solid curves).}
  \label{fig:newcorr}
\end{figure}

We calculate both the self and cross correlations between biases for four interpolation schemes, as shown in Fig.~\ref{fig:newcorr}. We find that both $C_c(\theta)$ and $C_m(\theta)$ are largely suppressed in cross correlations, especially in short-range and mid-range. In the rest of this section, our results are shown using cross correlation between different exposures.

\begin{figure*}[htbp]
\centering
\begin{minipage}{0.49\textwidth}
\centering
\includegraphics[width=8.1cm]{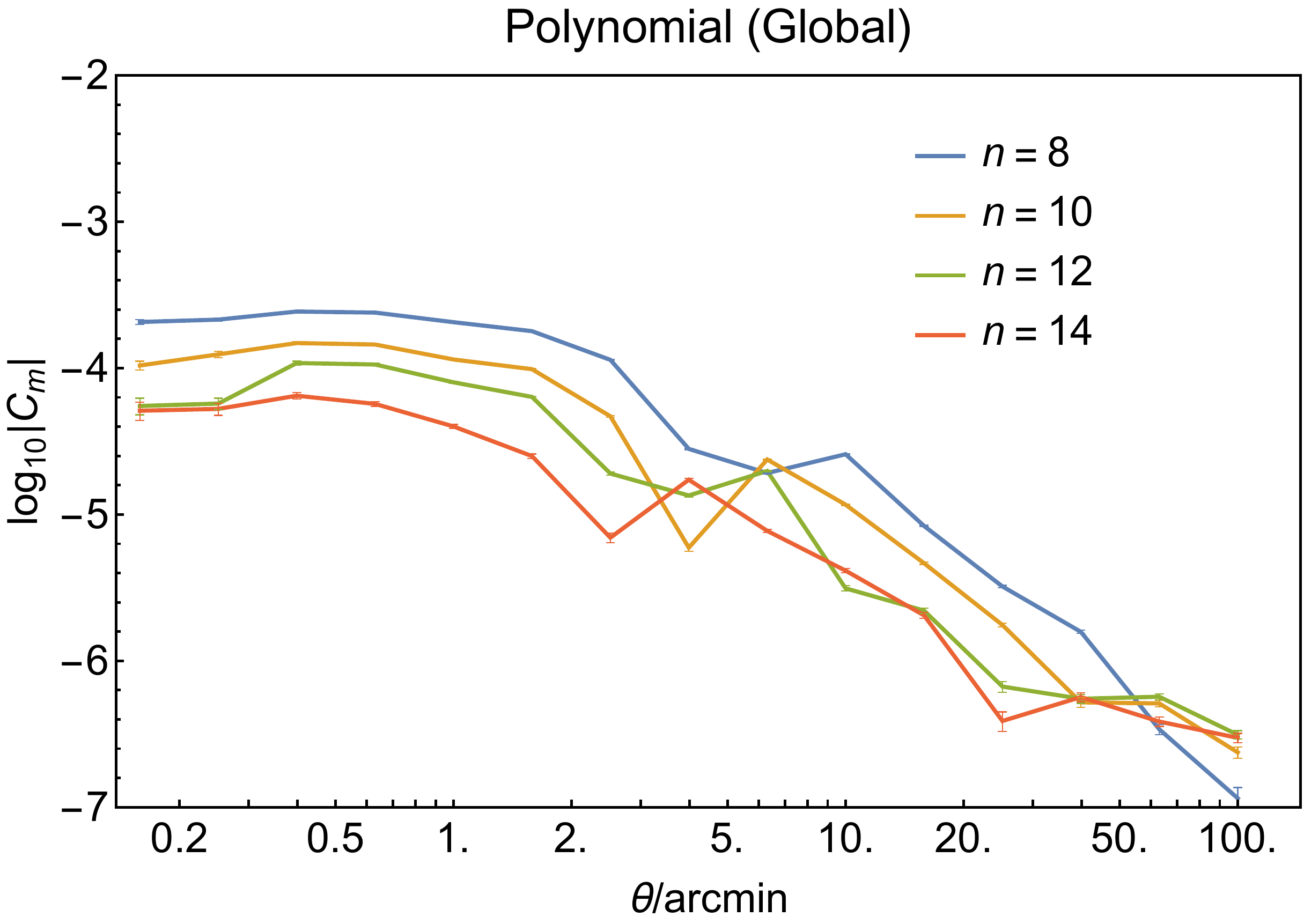}
\includegraphics[width=8.1cm]{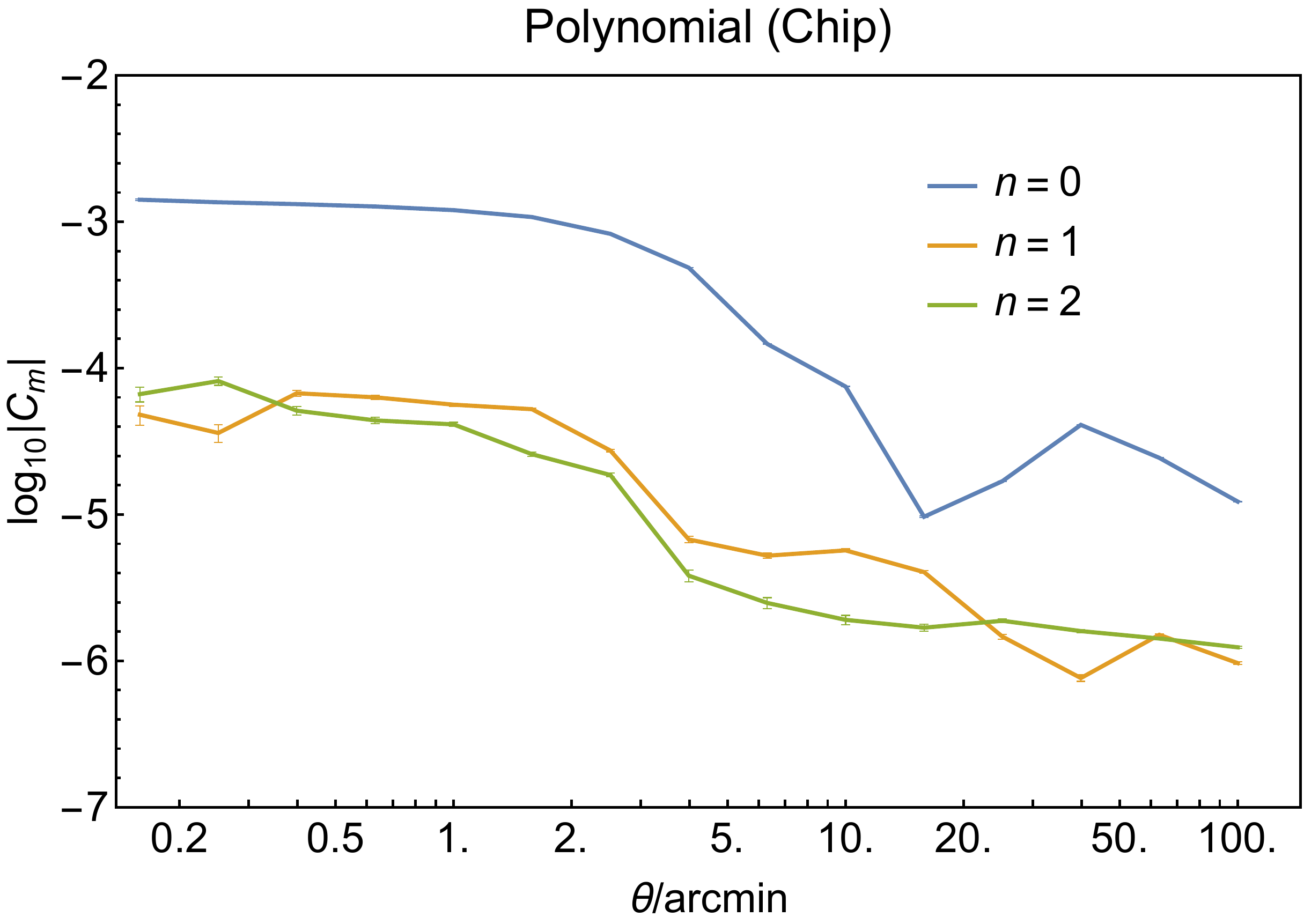}
\includegraphics[width=8.1cm]{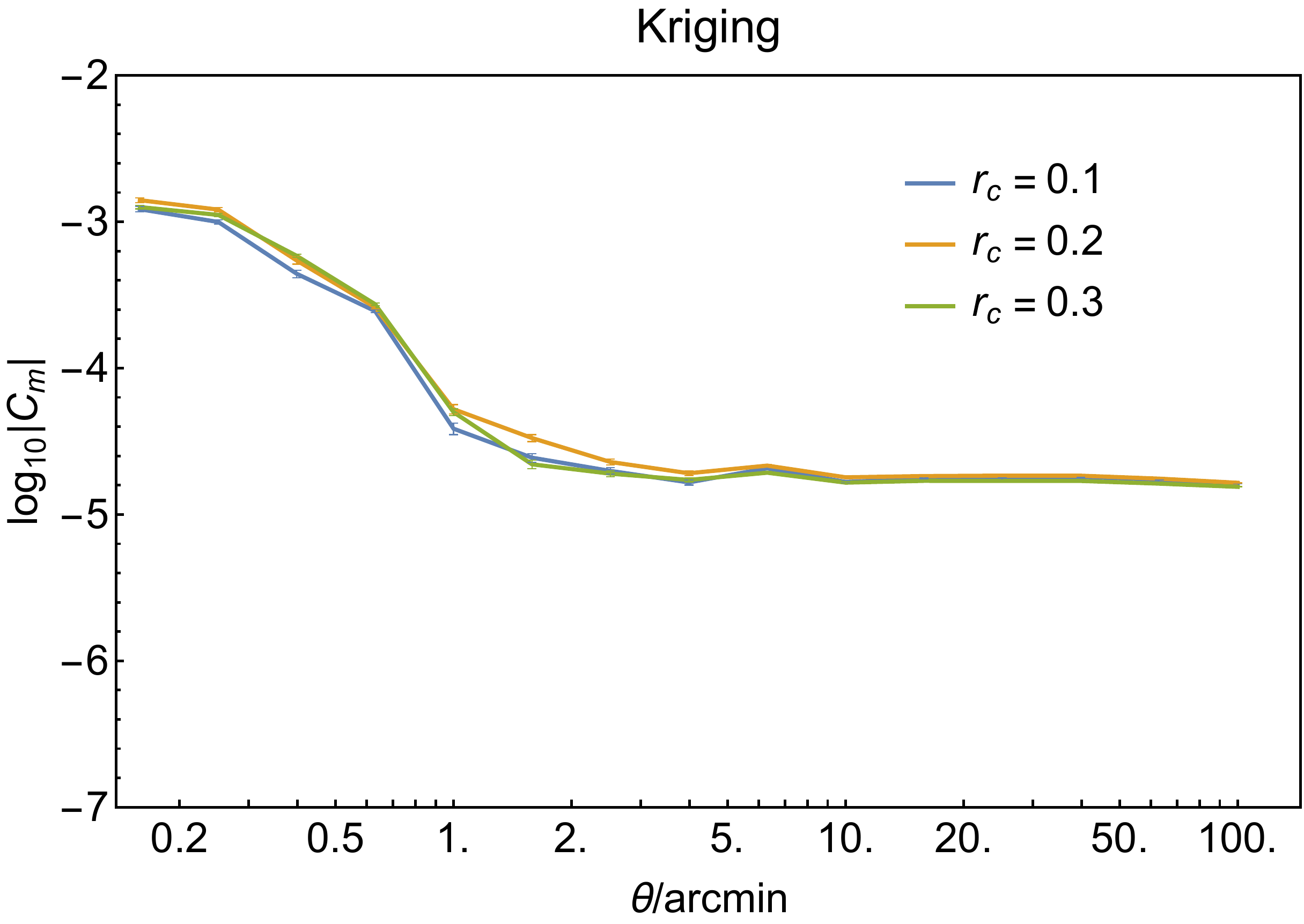}
\includegraphics[width=8.1cm]{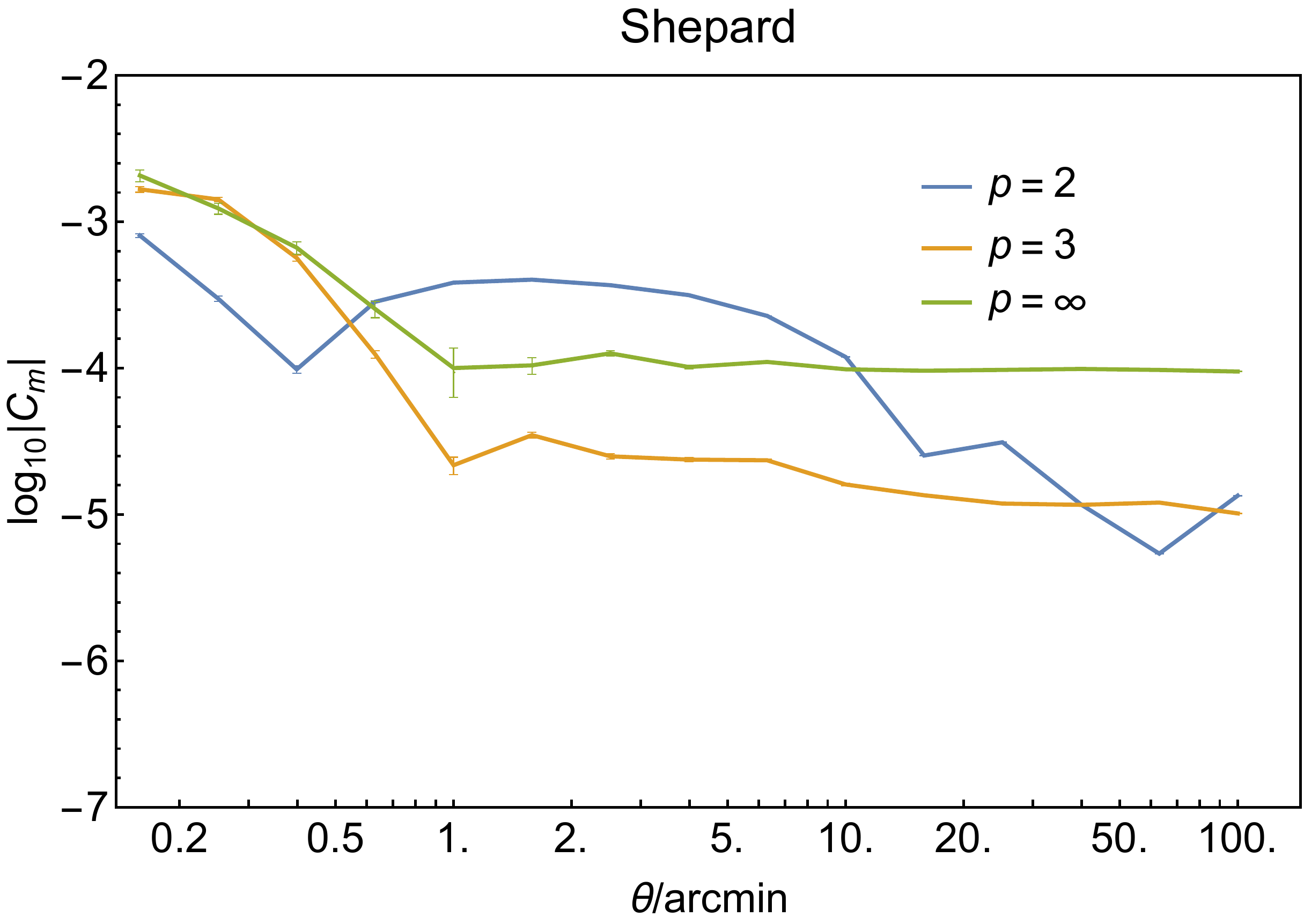}
\end{minipage}
\begin{minipage}{0.49\textwidth}
\centering
\includegraphics[width=8.1cm]{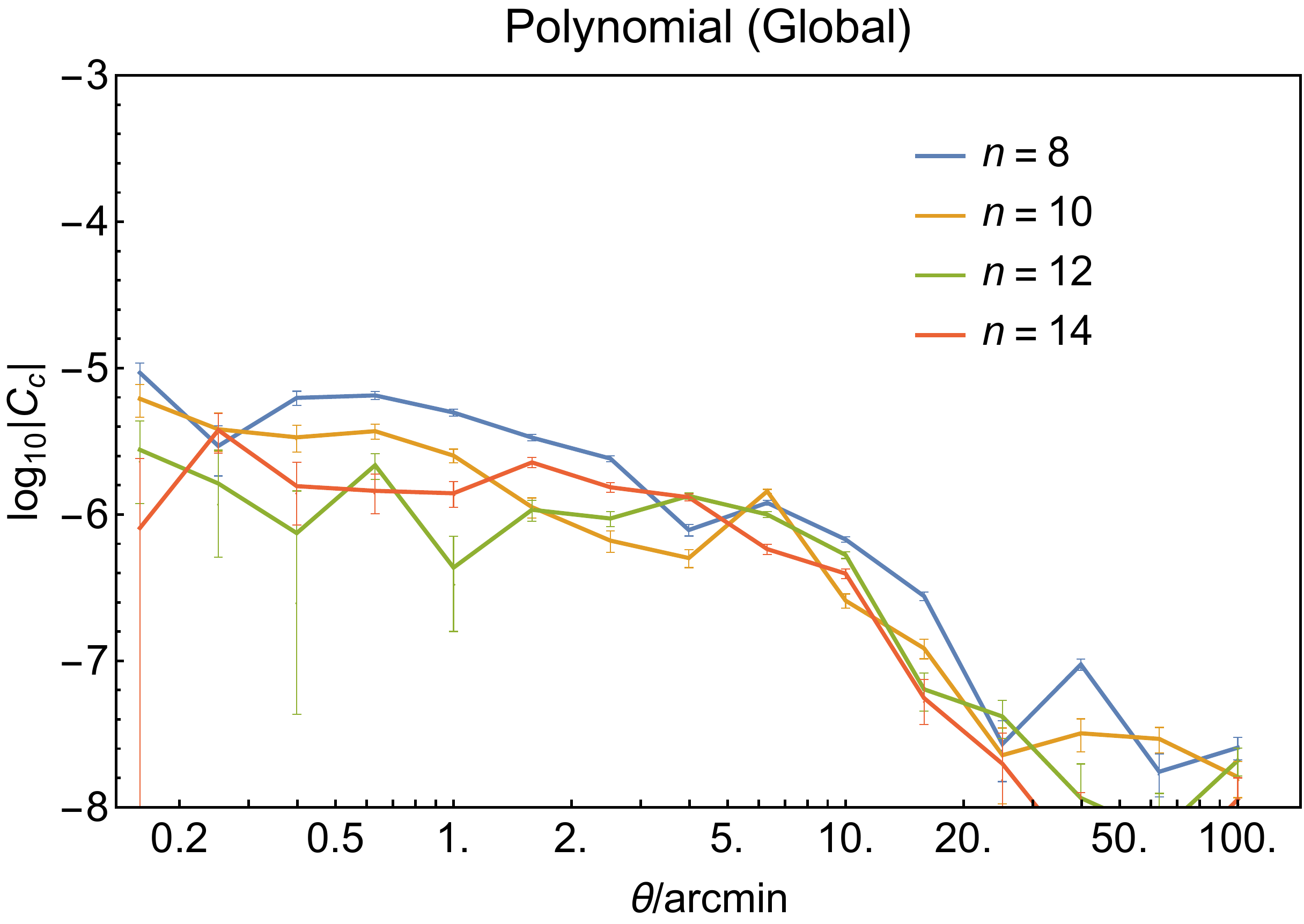}
\includegraphics[width=8.1cm]{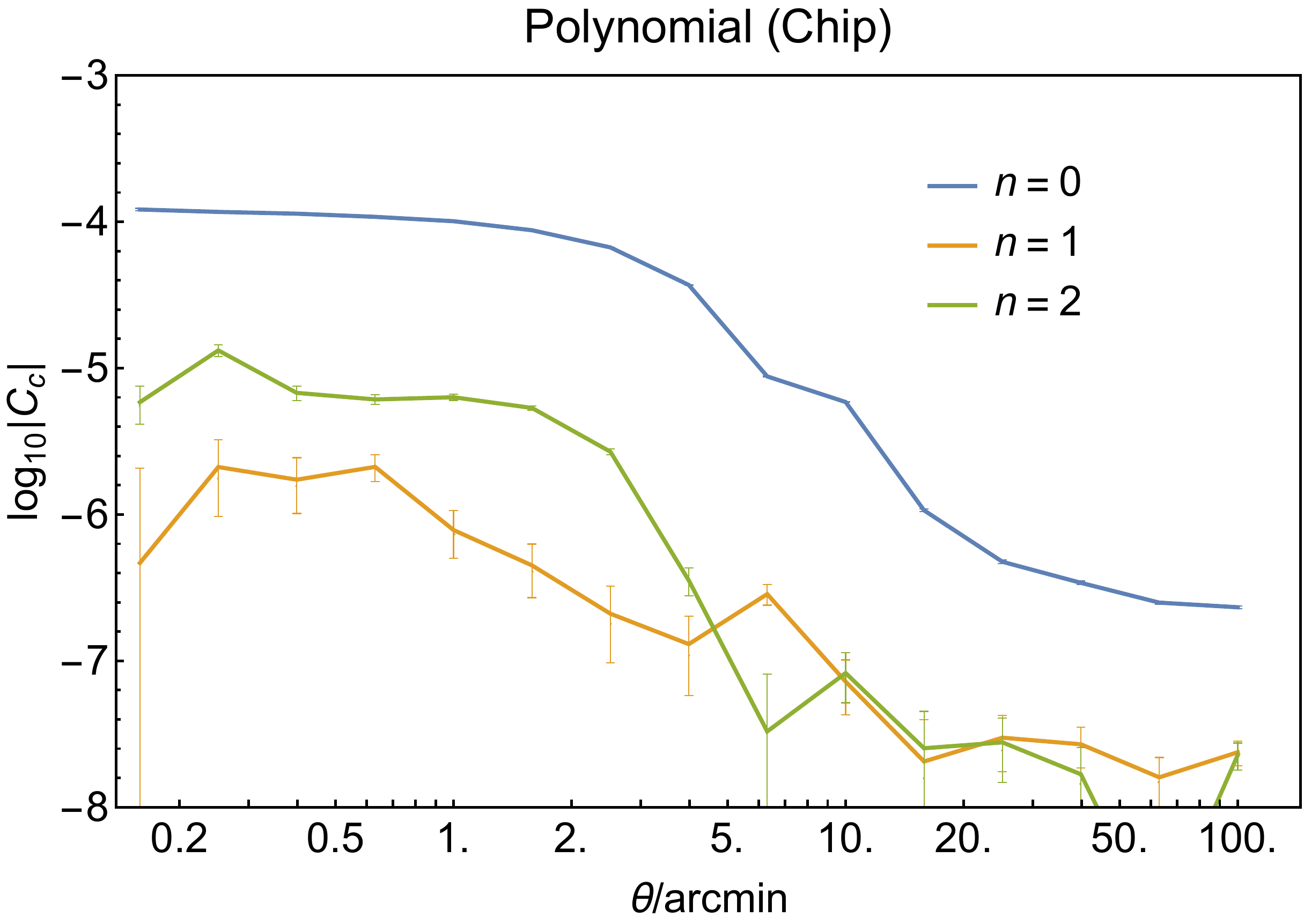}
\includegraphics[width=8.1cm]{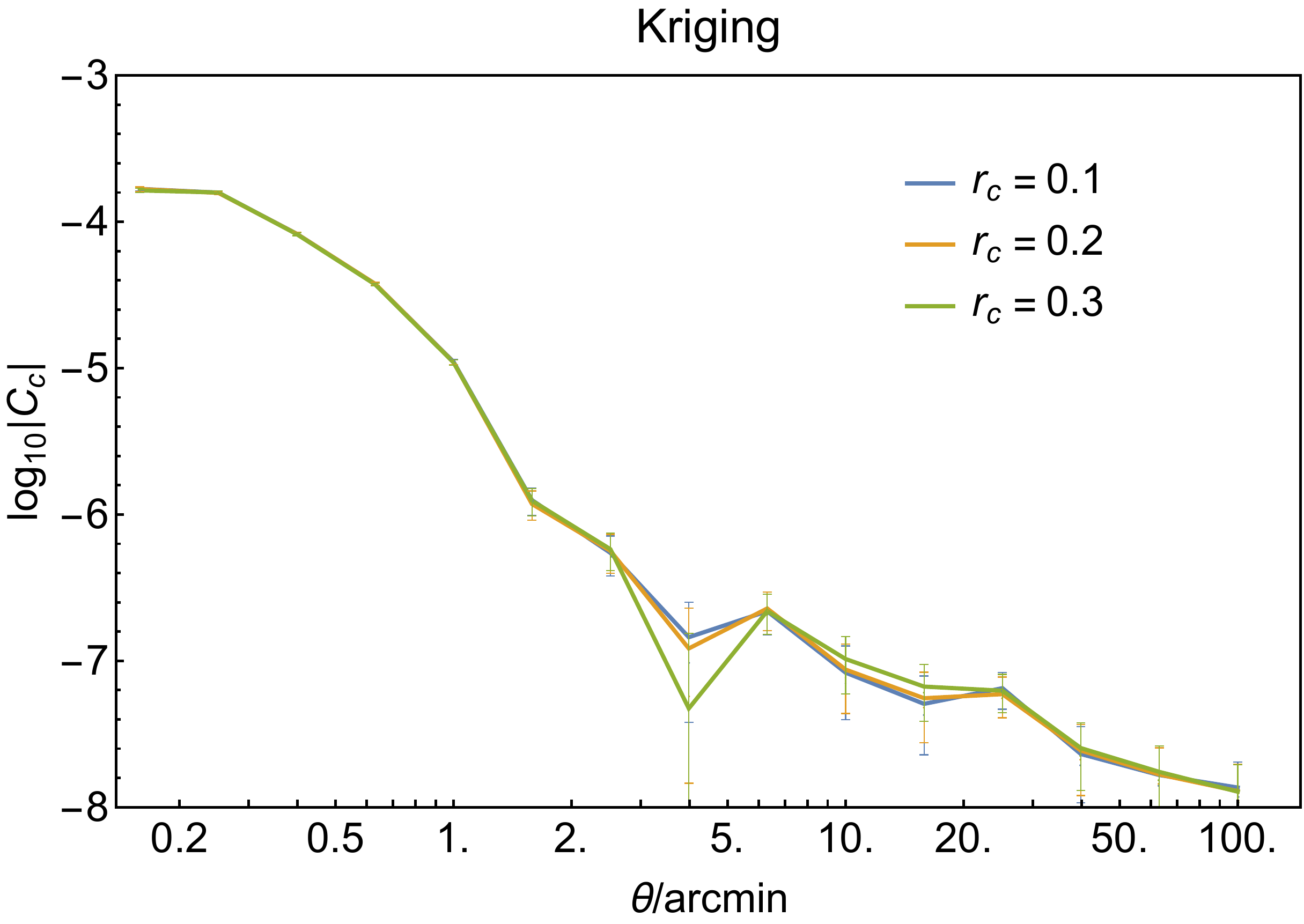}
\includegraphics[width=8.1cm]{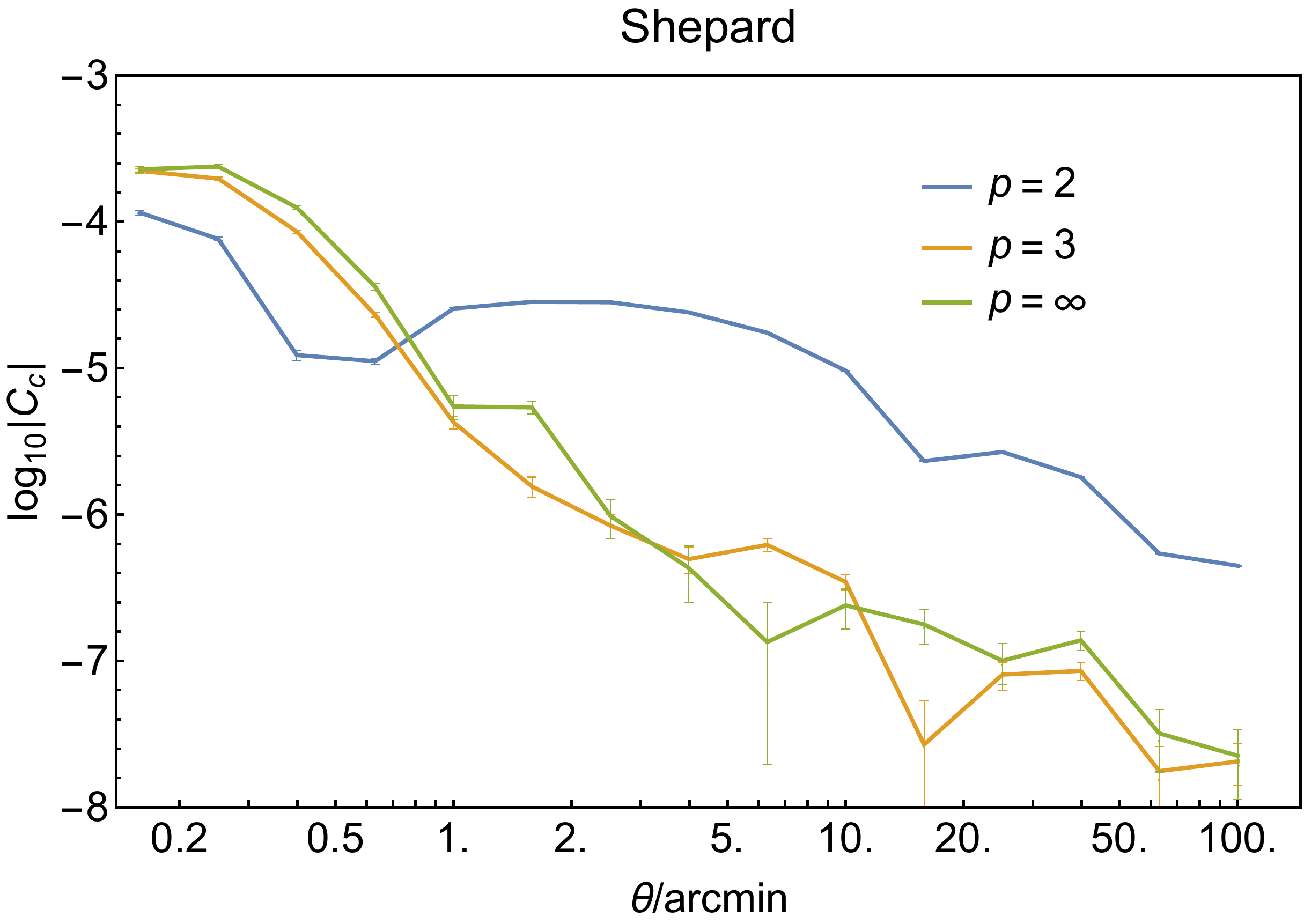}
\end{minipage}
  \caption{Correlation functions of $m$ and $c$ from different interpolation methods. The best implementation in each algorithm is: 1. 1st order chipwise polynomial fitting; 2. 10th order global polynomial fitting; 3. $r_c=0.3$ Kriging interpolation; 4. $p=3$ Shepard interpolation.}
  \label{fig:allcorr}
\end{figure*}

Comparisions between different implementations of the interpolation schemes are shown in Fig.~\ref{fig:allcorr}. In global polynomial fitting, we find that the performance is somewhat stablized when the order of the polynomial function $n$ is $\gsim 10$. In the case of chipwise polynomial fitting, $n=1$ generally works better than other choices. In the Kriging method, the value of $r_c$ does not seem to affect the correlations of the biases much. For the Shepard method, $p=3$ yields the best results. Combining the results of the previous sections, we can conclude that the best implementations of the interpolation schemes are: 1. 1st order chipwise polynomial fitting; 2. 10th order global polynomial fitting; 3. $r_c=0.3$ Kriging interpolation; 4. $p=3$ Shepard interpolation.

\begin{figure}[htbp]
  \includegraphics[width=8cm]{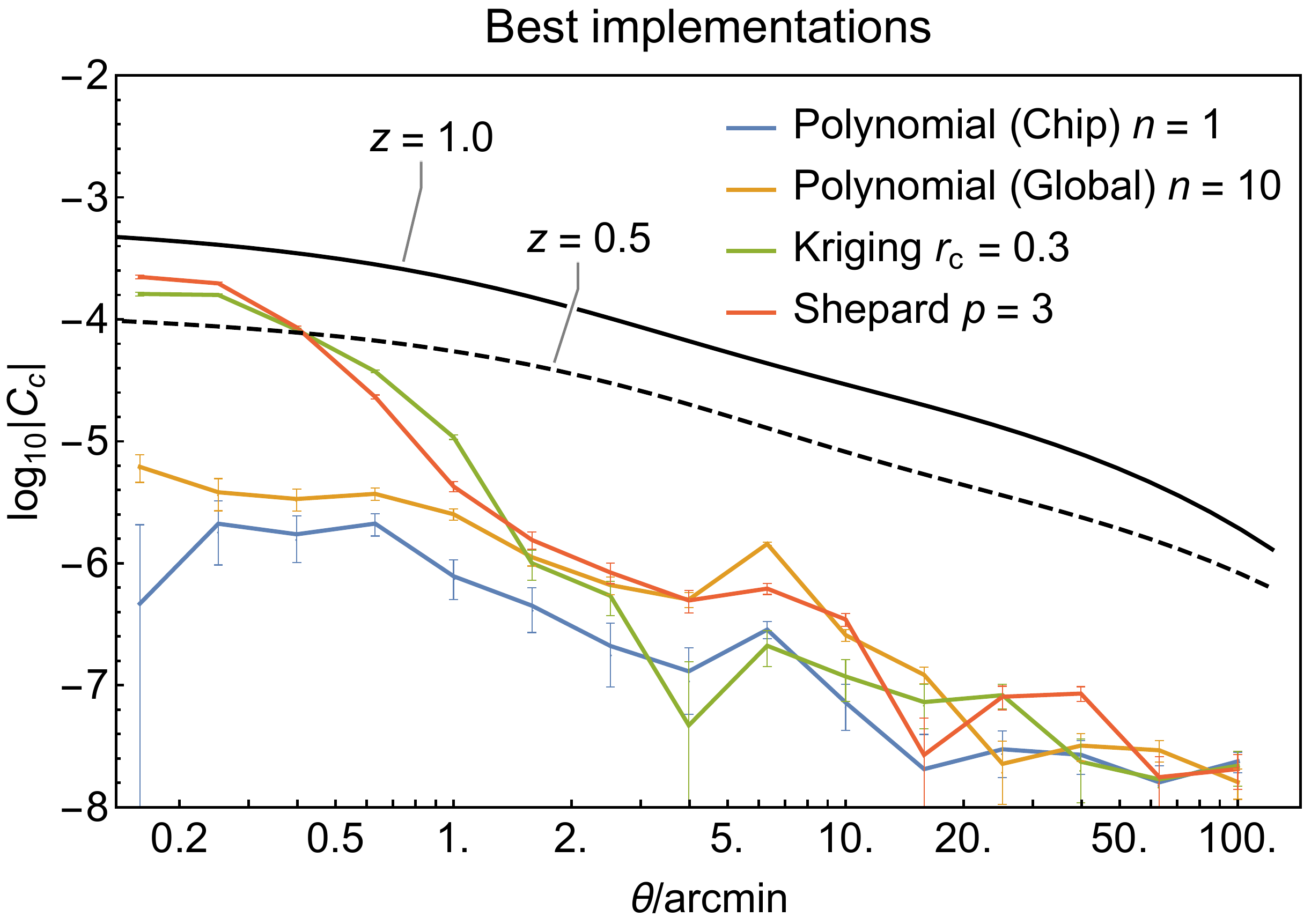}
  \caption{Correlation functions of $c$ from four best PSF interpolation methods. The black solid and dashed curves show the corresponding shear-shear correlations at redshift 1 and 0.5 respectively, predicted in $\Lambda$CDM model with $\Omega_m=0.283, \Omega_{\Lambda}=1-\Omega_m,\sigma_8=0.814, h=0.693$.}
  \label{fig:bestcorr}
\end{figure}

The comparisons of correlation functions between best implementations are shown in Fig.~\ref{fig:bestcorr}, in which we also show the corresponding shear-shear correlations ($\xi_+$) at redshifts 1 and 0.5 in the $\Lambda$CDM cosmological model with $\Omega_m=0.283, \Omega_{\Lambda}=1-\Omega_m,\sigma_8=0.814, h=0.693$. The parameters are derived with CFTHLenS+WMAP7+BOSS+R09 in \cite{kilbinger2013}, and the nonlinear density power spectrum is derived using the halofit code \citep{smith03}. We may divide the separation $\theta$ into three ranges: short-range ($\theta\lesssim1\ \mathrm{arcmin}$), mid-range ($1\ \mathrm{arcmin}\lesssim\theta\lesssim20\ \mathrm{arcmin}$), and long-range ($\theta > 20\ \mathrm{arcmin}$). In short-range, two polynomial interpolations are much better (generally 10 to 100 times) than Kriging and Shepard. In mid-range, four methods perform similarly well. In long-range, two polynomial interpolations work better than Kriging and Shepard, especially in terms of $C_m$. Note that the constant plateau in the $C_m$ of Kriging or Shepard on large scales is caused by $\langle m \rangle$, which is not subtracted in calculating $C_m$. Within the two kinds of polynomial fitting methods, the chipwise fitting is a little bit better than the global fitting. We can conclude that overall, the best interpolating scheme is the 1st order chipwise polynomial fitting. The global polynomial fitting of 10th-order is also competitive.

We can draw an analogy between the residual PSF ellipticities $(e_1, e_2)$ and the additive biases $(c_1, c_2)$ calculated in this paper. We find that the amplitudes of our correlation functions $C_c$ are roughly consistent with $\xi_+$ measured previously in \cite{h12}, assuming that the residual PSF errors are mainly due to atmospheric turbulence, and the galaxy sizes are comparable to the PSF sizes. Note that we need to rescale the correlation according to the exposure time (600 seconds each) of the image we use \citep{dev07}. 

\section{Discussion}
\label{sec:discussion}

\subsection{Sensitivity to outliers}
\label{sec:outliers}

In real data, outliers are not only possible but also regular due to a few reasons: binary stars, star-galaxy overlap, cosmic rays atop stars, and CCD defects. There is no doubt that outliers have observable impacts on the quality of interpolation, but the degrees are not necessarily the same in different schemes.

Since the polynomial schemes is fitted on a certain domain (at least the area of a chip), the impact of outlier is not significant locally, but it is carried to distant positions. On the contrary, Kriging and Shepard are sensitive to outliers more locally, which are screened by other stars rapidly when the distance increases. Such contrast can easily be verified by Fig.~\ref{fig:bestcorr}, where polynomial performs much better than Kriging and Shepard at short-range, while such advantage rapidly diminishes as the range goes longer.

It is generally believed that Kriging is better at reflecting small scale details than other schemes like polynomial \citep{berge2012}. However, this feature might not turn out to be an advantage, because it makes Kriging more prone to outliers. The impact of outliers in polynomial schemes is not as dominant on small scale as they are in Kriging and Shepard interpolation, and on large scales, the impacts are likely cancelled out by other outliers at a certain level.

Here are a few other points regarding the validity of the conclusions of \cite{berge2012} on real data: 1. In their simulations, there are no real outliers, as all the stars used for PSF reconstruction are well defined; 2. The spatial distributions of the PC (Principle Component) weights, which are used to form the PSF morphology at each star location, are generated as Gaussian fields, conforming to the condition for the best performance of Kriging by definition; 3. Fig.~2, 3, and 4 of \cite{berge2012} seem to indicate that one can further improve the performance of polynomial fitting by reducing the domain size.

To better understand the effect of outliers, we try a reasonable recipe to remove outliers in the ``reconstruction'' group. The policy is as follows:

\begin{enumerate}
  \item use chipwise polynomial of 1st order to interpolate the stars in the ``reconstruction'' group based on themselves;
  \item the biases $\{m_1, c_1, m_2, c_2\}$ of the reconstructions are measured, and the mean values and standard deviations within each exposure are calculated;
  \item stars with very large biases, i.e. any of $\{m_1, c_1, m_2, c_2\}$ being out of the corresponding $3\sigma$ range, are removed from the ``reconstruction'' group.
\end{enumerate}

\begin{figure}[htbp]
  \includegraphics[width=8cm]{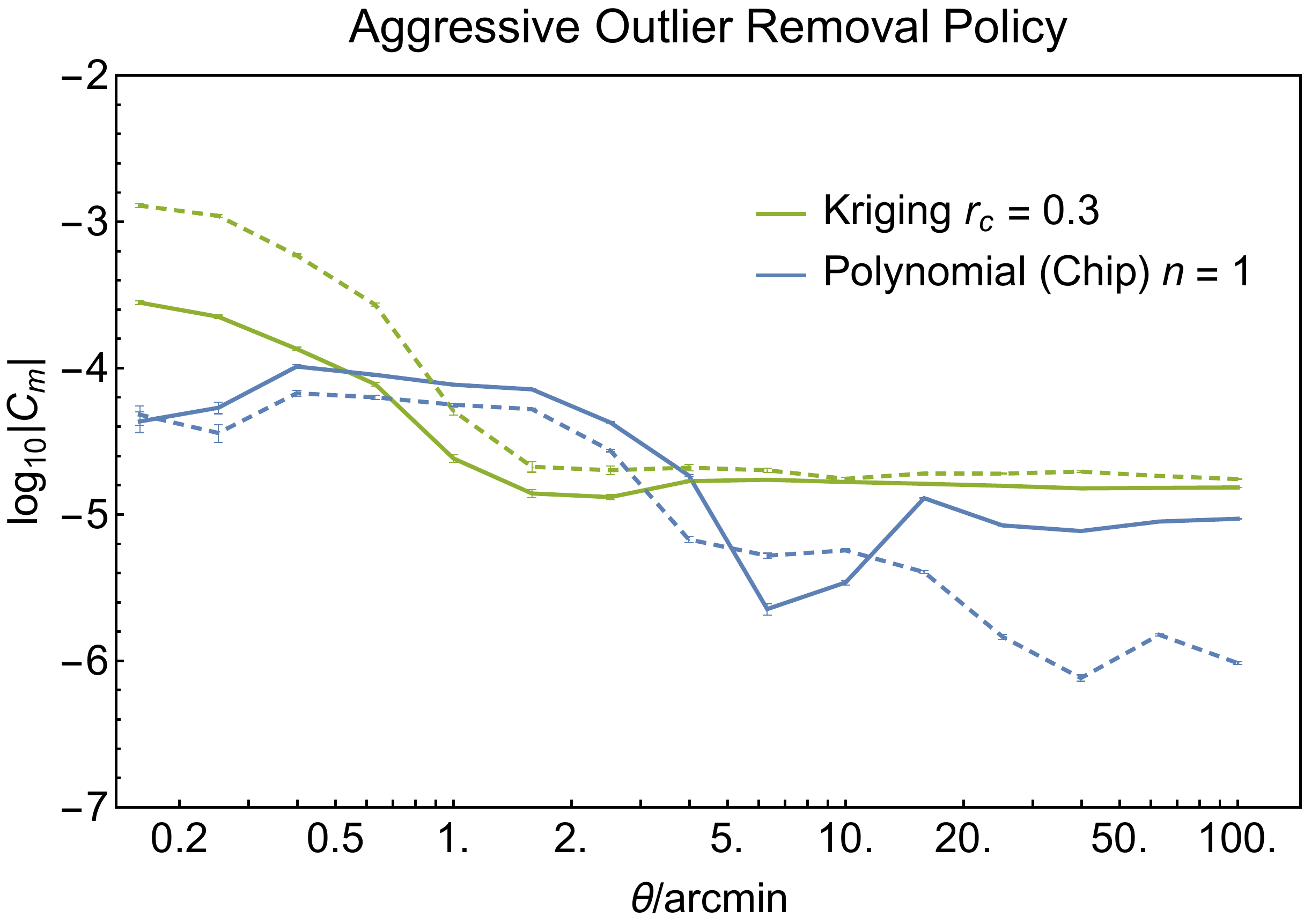}
  \includegraphics[width=8cm]{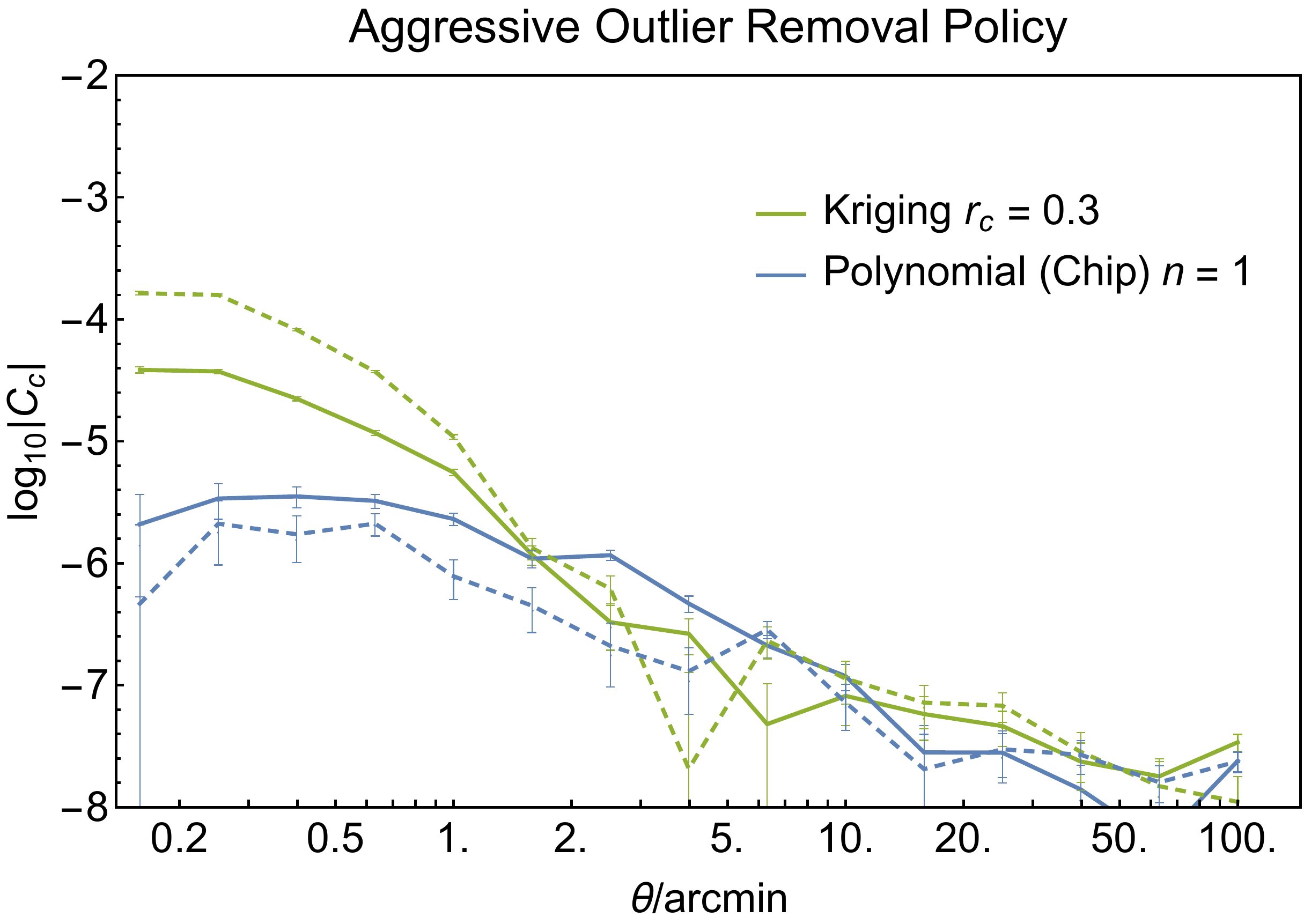}
  \caption{The spatial correlation functions of the additive and multiplicative shear biases calculated from the PSF's reconstructed with (the dashed curves) or without (the solid curves) the ``outliers'' in the "reconstruction" group of stars.}
  \label{fig:remove}
\end{figure}

\begin{figure}[htbp]
  \includegraphics[width=8cm]{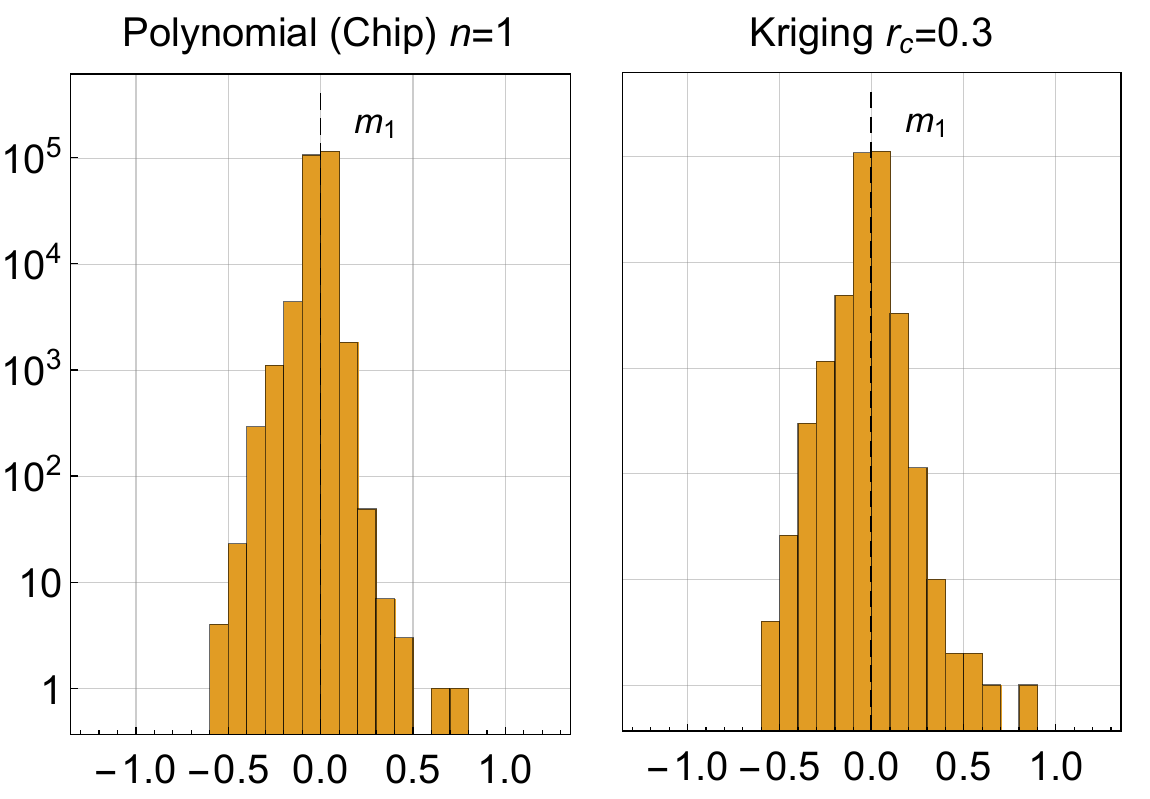}
  \caption{The distributions of the multiplicative shear biases calculated from the PSF's reconstructed without the ``outliers'' in the "reconstruction" group of stars.}
  \label{fig:remove2}
\end{figure}

The performances of Kriging and chipwise polynomial fitting adopting the above policy are compared with their original ones in Fig.~\ref{fig:remove}, in which the solid curves are the new results, and the dashed curves are the old ones. We find that after removing the so-called outliers, the short-range performance of Kriging is indeed improved. It also leads to the suppression of the long tail in the distribution of the multiplicative shear bias in Kriging (but not much change for polynomial fitting), as shown in Fig.~\ref{fig:remove2}. However, on the other hand, the long-range performance of the chipwise polynomial fitting becomes somewhat worse. The later is possibly an indication of certain selection bias introduced by our outlier removal. One therefore has to be careful with outlier removal. A further study of this topic is beyond the scope of this paper.

\subsection{Overfitting}
\label{sec:overfitting}

\begin{figure}[htbp]
\centering
\includegraphics[width=8cm]{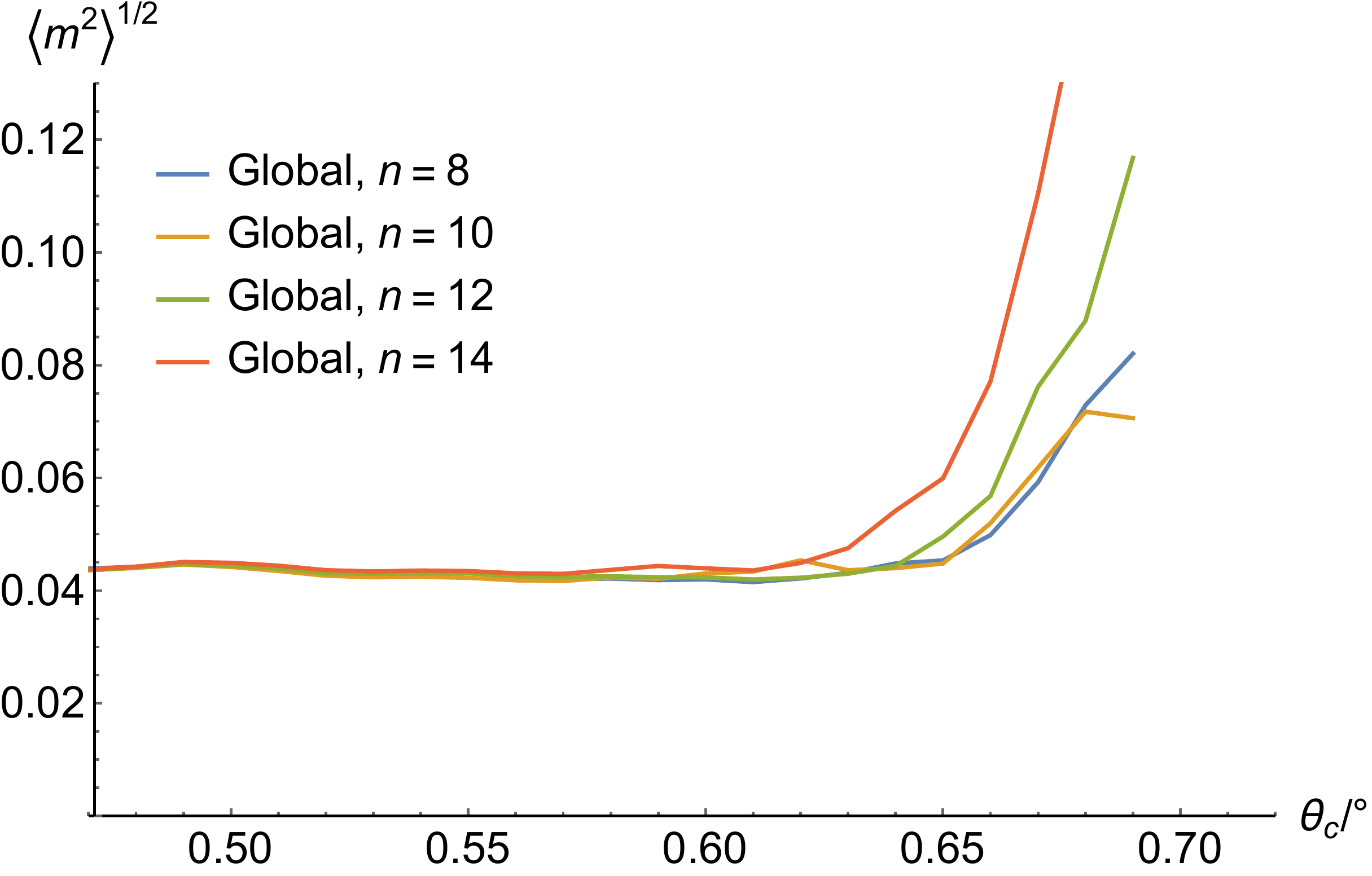}
\caption{The variance of $m$ as a function of $\theta_\mathrm{c}$, the distance from the interpolated sites to the center of their exposures. The maximum distance is about $0.7^\circ$ -- the distance from the center to the corner of a $1^\circ \times 1^\circ$ exposure.}
\label{fig:overfitting}
\end{figure}

Polynomial interpolation implicitly assumes that the spatial change of the PSF morphological properties (or pixel values) can be modelled by a 2D polynomial function. In the case of global polynomial fitting, the assumption is somewhat too restrictive. To investigate this phenomenon further, we plot $\langle m^2 \rangle^{1/2}$ against the distance from the center of the exposure in Fig.~\ref{fig:overfitting}. We find that the variance of $m$ increases when the order of the fitting function keeps increasing above 10, implying that overfitting starts to become important. This fact is most easily seen at the edge of the fitting domain. Normally, interpolation at the corners of a domain is more difficult. At this point, it is not clear whether a different functional form for global fitting (\eg, see \cite{hoekstra2004}) can further improve the PSF reconstruction accuracy. More options may be tried in a future work.

It is interesting to point out that in the chipwise fitting case, there is almost no room to further improve the fitting, as the 2nd order function already shows certain level of overfitting. Therefore, the residual PSF errors in the chipwise case are mainly caused by the stochastic component of PSF spatial variation, rather than the choice of the functional form. The stochastic component simply cannot be easily modelled by smooth functions.

\subsection{Effect of stellar number density}

\begin{figure}[htbp]
  \includegraphics[width=8cm]{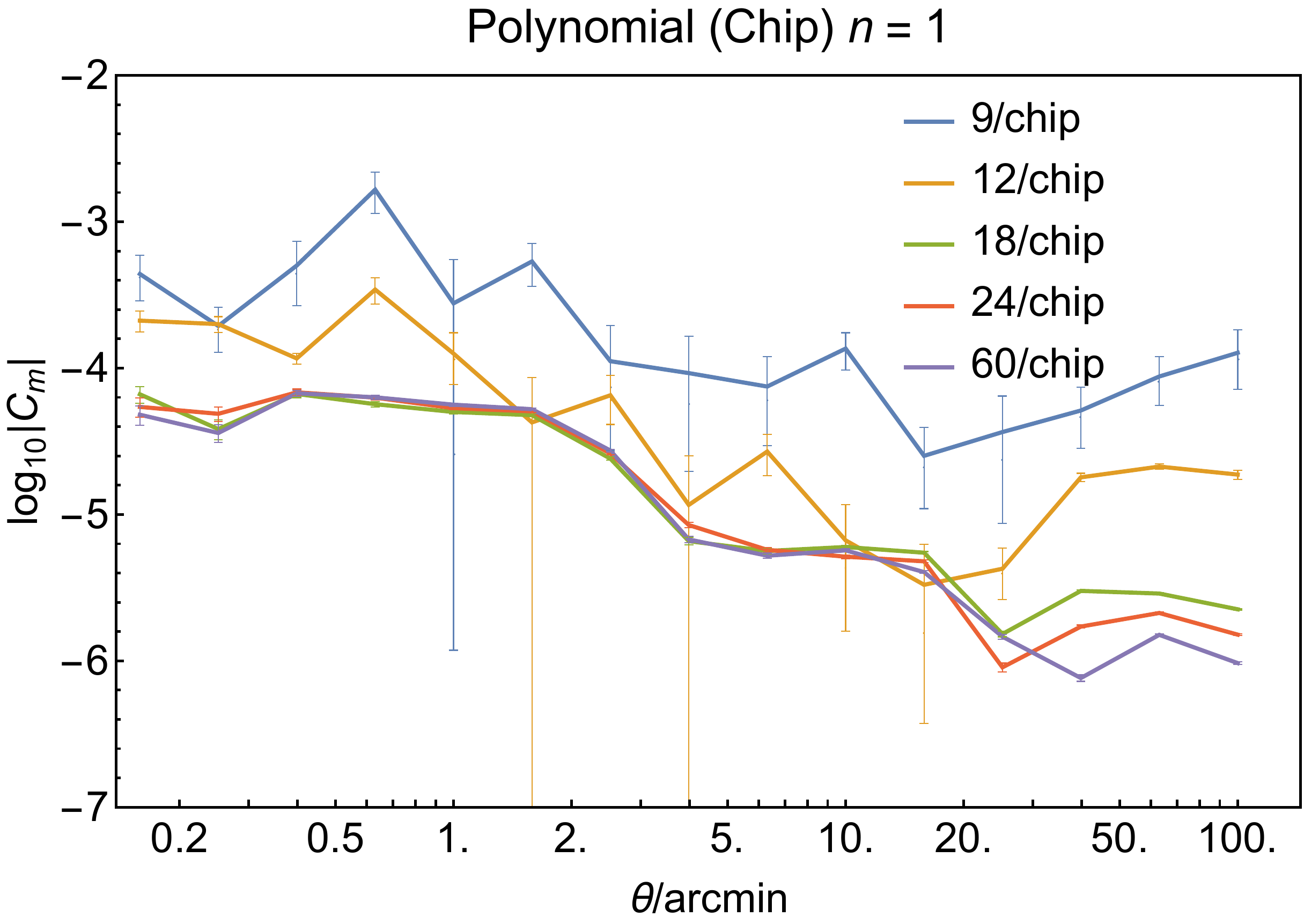}
  \includegraphics[width=8cm]{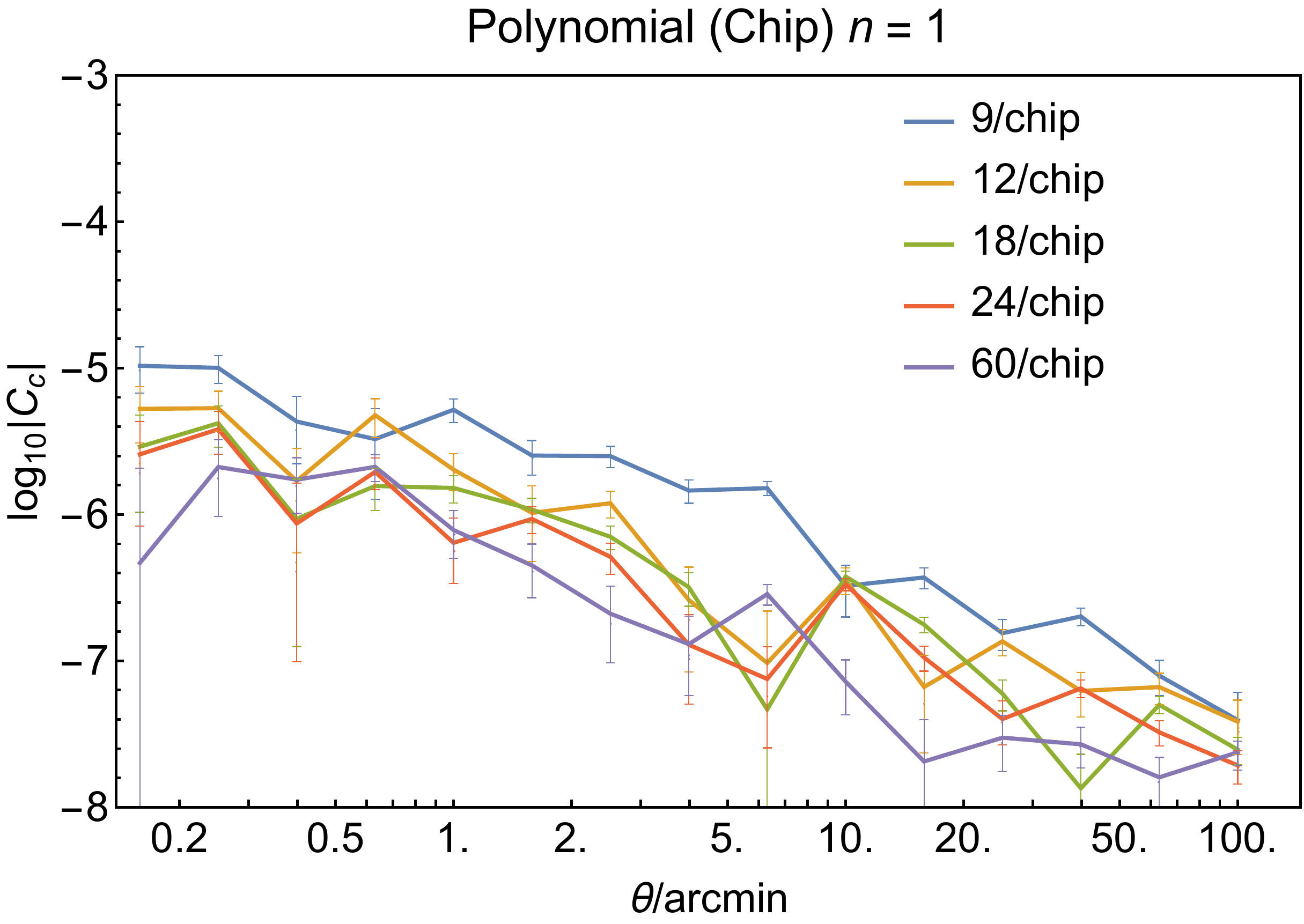}
  \caption{Comparison between the performances of the 1st-order chipwise polynomial fitting with different stellar number density.}
  \label{fig:density}
\end{figure}

It is well known that the number density of stars affects the accuracy of PSF reconstruction. Too few available stars cannot give adequate information about the PSF distribution.  We investigate the minimum requirement for stellar number density. In our previous tests, on average, half of all available stars ($\gsim 50$) per chip are used to do interpolation. Here we cut down the number density to as low as 9 stars per chip. The results are shown in Fig.~\ref{fig:density}. We find that the improvement of performance is negligible when the stellar number density is higher than 20 per chip, corresponding to about $0.2 {\rm stars /arcmin^2}$ (with ${\rm SNR}\geq 100$). Further increasing the stellar number density in the reconstruction group does not lead to any significant improvement. This requirement is satisfied in all exposures of the CFHTlenS data.

\subsection{Alternative galaxy generation methods}

It is useful to test the performance of PSF interpolation with different galaxy types to make our conclusion more general. We try generating galaxies of larger and smaller average sizes using gaussian distribution, corresponding to the standard deviations of the point source positions $\sigma$ changing from $1.4\ \mathrm{pixel}$ to $4.2\ \mathrm{pixel}$ or $0.8\ \mathrm{pixel}$. We also try generating point sources in galaxies using random walks \citep{jz08} instead of Gaussian distribution. In this method, each galaxy is also composed of 20 point sources of equal luminosity, the position of which are determined by the end points of random walk steps with step size $l = 1.0\ \mathrm{pixel}$. To limit the size of the galaxy, the random walks restart from the center once its distance to the center is greater than $d_\mathrm{max} = 8\ \mathrm{pixel}$. The resulting galaxies have rms sizes similar to those of the Gaussian ones with $\sigma=1.4\ \mathrm{pixel}$. 

\begin{figure}[htbp]
  \centering
  \includegraphics[width=8cm]{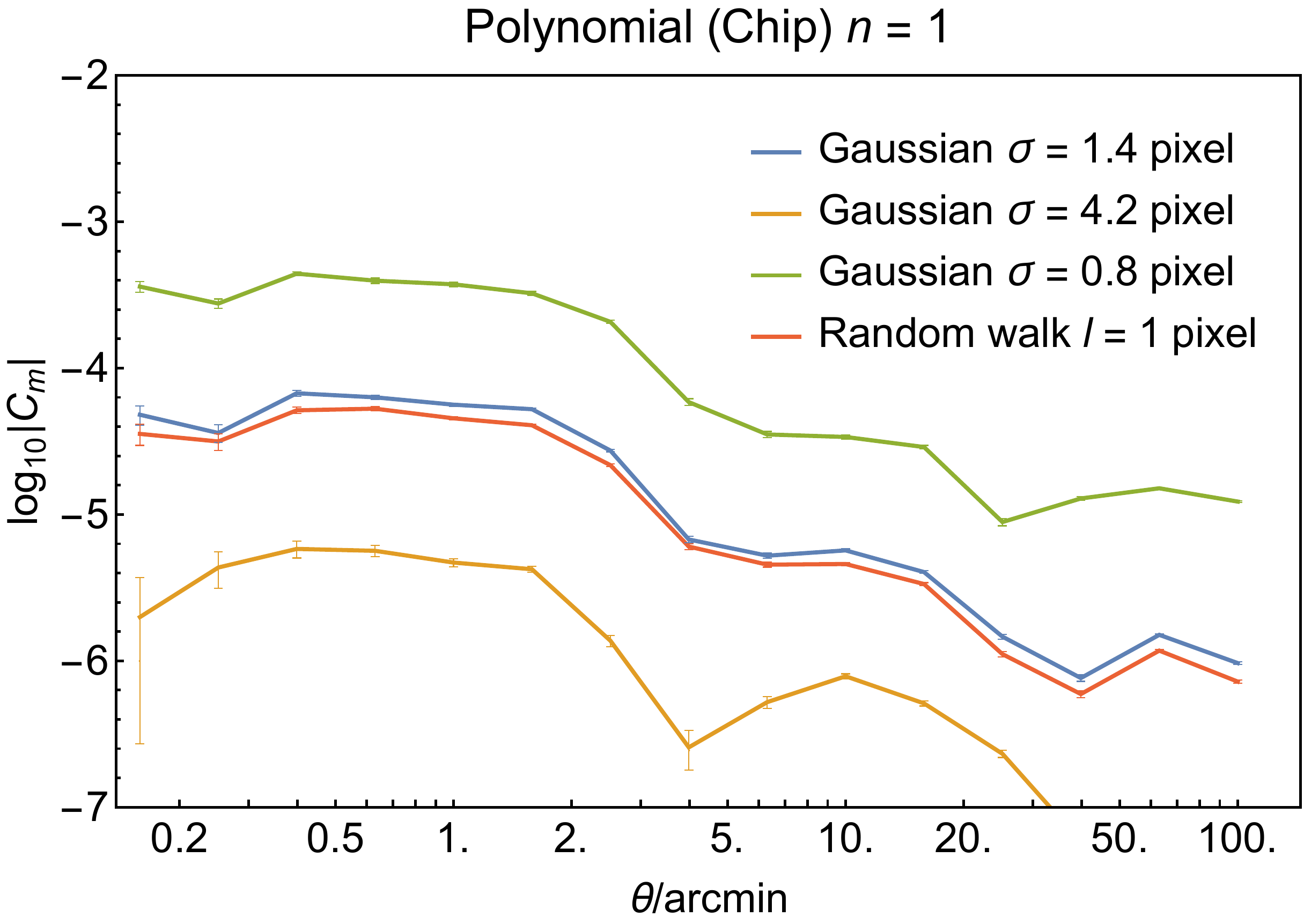}
  \includegraphics[width=8cm]{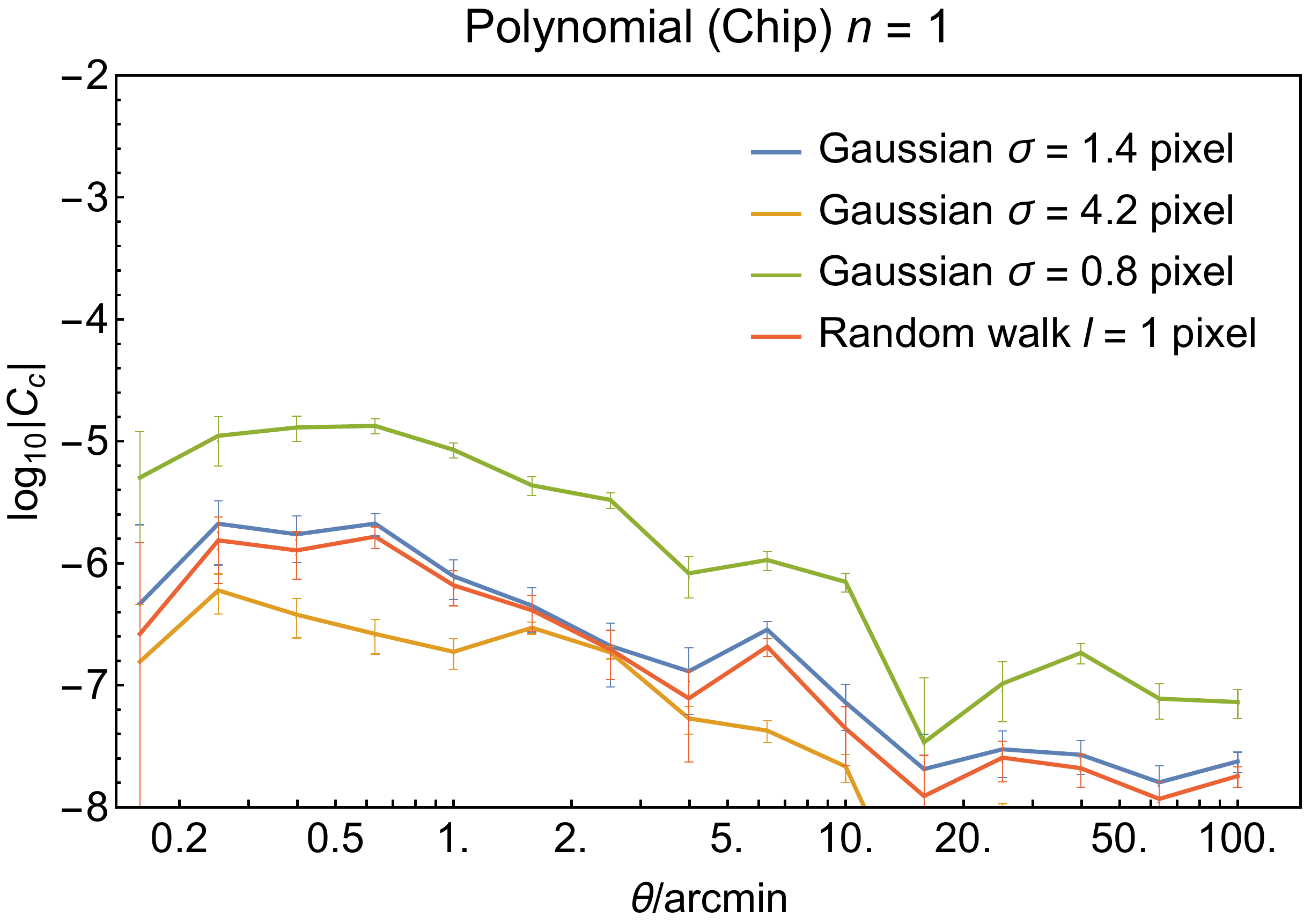}
  \caption{Comparison between the performances of the 1st-order chipwise polynomial fitting with four different galaxy generation methods.}
  \label{fig:bestcorrrw}
\end{figure}

The resulting correlation functions of the shear biases are shown in Fig.~\ref{fig:bestcorrrw}, which indicates that the impact of PSF error on shear recovery is closely related to the galaxy size rather than the galaxy morphology. This point is expected in ZLF15 (but not necessarily true in other shear measurement methods). In general, larger galaxies are less influenced by the PSF, and the shear measurements are less sensitive to the accuracy of PSF reconstruction.

\section{Conclusion}
\label{sec:conclusion}

Accurate reconstruction of the PSF form at the galaxy position is a crucial step in cosmic shear measurement. This is typically done by interpolating the neighboring star images. We have developed a pipeline based on real data (real star images and spatial distributions) to test the performances of several popular interpolation schemes, including polynomial fitting, Kriging, and Shepard. The pipeline does not only check the quality of the reconstructed PSF field itself, but also allows one to directly test the impact of PSF reconstruction error on shear recovery accuracy within any specific shear measurement method. The latter is achieved by generating a large number of simulated galaxy images at the position of real stars.

Applying our pipeline on the CFHTlenS data, we can draw several conclusions so far:
\begin{itemize}
  \item The chipwise 1st-order polynomial fitting and the exposure-wide 10th-order polynomial fitting consistently yield better reconstruction results than the other schemes tested;
  \item The shear-shear correlation bias caused by the spatially correlated residual PSF errors can be significantly reduced by correlating the shear fields from two different exposures;
  \item In the two best schemes, the spatial correlations of the additive shear recovery biases induced by the residual PSF reconstruction error are about $10^{-7}-10^{-6}$ on the scale of $1-100$ arcmin for galaxies of pre-seeing sizes comparable to the PSF size, much less than the weak lensing signal at reasonably redshifts, but could be a serious concern for high precision shear-shear correlation measurement;
  \item The Kriging and Shepard methods consistently yield much higher correlations of the additive shear biases on small scales, mostly due to their oversensitiveness to outlier in the population of star images;
  \item To achieve the best performance in PSF reconstruction, the best schemes only require no more than 20 stars per chip (corresponding to about $0.2 {\rm stars /arcmin^2}$) with ${\rm SNR}\geq 100$ on average, a condition that is satisfied in all CFHTlenS fields. Lower stellar number density leads to worse PSF fitting and therefore larger shear recovery biases. Higher stellar number density does little in improving the performance of PSF interpolation accuracy in the methods of this paper;
  \item Our conclusions are not sensitive to the choice of the galaxy morphology (either Gaussian disks or random profiles determined by point sources that are connected by random walks), but strongly depend on the galaxy size: larger galaxy sizes lead to smaller shear biases due to PSF reconstruction error, and vice versa.
\end{itemize}

It is interesting to observe that in the case of chipwise polynomial fitting, even the 2nd-order fitting function yields an overfitting case. It suggests that the residuals in the case of 1st-order polynomial fitting cannot be modelled as a smooth distribution, and a better PSF modelling should introduce certain stochasticity. Some recently developed PSF interpolation schemes, such as the PSFENT method \citep{chang12}, may have the potential of further reducing the residual PSF errors. This direction deserves more investigations in the near future. More statistical tools such as the E- and B-mode aperture mass variances \citep{schneider98,schneider02,hamana13} will also be considered in our future work.

\acknowledgments{We would like to thank the referee for very useful suggestions for improving this paper. This work is based on observations obtained with MegaPrime/MegaCam, a joint project of CFHT and CEA/DAPNIA, at the Canada-France-Hawaii Telescope (CFHT) which is operated by the National Research Council (NRC) of Canada, the Institut National des Sciences de l\'Univers of the Centre National de la Recherche Scientifique (CNRS) of France, and the University of Hawaii. This research used the facilities of the Canadian Astronomy Data Centre operated by the National Research Council of Canada with the support of the Canadian Space Agency. CFHTLenS data processing was made possible thanks to significant computing support from the NSERC Research Tools and Instruments grant program.

The processing of single exposure images is conducted using the THELI software, a tool for the automated reduction of astronomical images developed by the CFHTlenS team \citep{erben2005,schirmer2013}.

JZ is supported by the NSFC grants (11673016 and 11433001), the National Key Basic Research Program of China (2013CB834900 and 2015CB857001), the National Thousand Talents Program for Distinguished Young Scholars, a grant (No.11DZ2260700) from the Office of Science and Technology in Shanghai Municipal Government. Dezi Liu and Zuhui Fan are supported by the NSFC grants (11333001 and 11173001), and by the Strategic Priority Research Program “The Emergence of Cosmological Structures” of the Chinese Academy of Sciences Grant No. XDB09000000. L.P.F. acknowledges the support from NSFC grants (11333001 and 11673018), STCSM grants 13JC1404400 and 16R1424800, and SHNU grant DYL201603. GL is supported by the One-Hundred-Talent fellowships of CAS，the NSFC grants (11273061 and 11333008)，National Key Basic Research Program of China (2015CB857000), and the ‘Strategic Priority Research Program the Emergence of Cosmological Structures’ of the Chinese Academy of Sciences (XDB09000000).

}

\end{document}